\documentclass{JHEP}
\usepackage{epsfig}
\usepackage{amssymb}

\newcommand{\ba}{\begin{array}}
\newcommand{\ea}{\end{array}}
\newcommand{\bt}{\begin{tabular}}
\newcommand{\et}{\end{tabular}}
\newcommand{\be}{\begin{equation}}
\newcommand{\ee}{\end{equation}}
\newcommand{\bea}{\begin{eqnarray}}
\newcommand{\eea}{\end{eqnarray}}
\newcommand{\bean}{\begin{eqnarray*}}
\newcommand{\eean}{\end{eqnarray*}}
\newcommand{\mat}[1]{\left( \matrix{#1} \right)}

\def\bX{\mathbf X}
\def\Y{\mathbf Y}

\def\vol{\mbox{vol}}
\def\Vol{\mbox{Vol}}
\def\IZ{\mathbb{Z}}
\def\IR{\mathbb{R}}

\def\IC{\mathbb{C}}
\def\IP{\mathbb{P}}

\def\cM{{\mathcal M}}

\def\beq{\begin{equation}}
\def\eeq{\end{equation}}

\def\Tr{\mathop{\rm Tr}}
\newcommand{\fref}[1]{Figure~\ref{#1}}
\def\eref#1{(\ref{#1})}
\def\sref#1{\S~\ref{#1}}

\def\nn{\nonumber}
\def\sign{\mbox{sign}}
\preprint{MIT-CTP-3473\\ UPR-1065-T\\NSF-KITP-04-23\\{hep-th/0402120}}

\def\cale{{\mathcal E}}
\def\calh{{\mathcal H}}
\def\cali{{\mathcal I}}
\def\Rt{\widetilde{R}}

\title{Chaotic Duality in String Theory}
\author{Sebasti\'{a}n Franco$^1$, Yang-Hui He$^2$, 
Christopher Herzog$^3$ and Johannes Walcher$^3$\\
\vskip .1cm
1. \parbox[t]{6in}{Center for Theoretical Physics,
Massachusetts Institute of Technology\\[-.1cm]
Cambridge, MA 02139, USA}\\[.2cm]
2. \parbox[t]{6in}{Department of Physics and Math/Physics RG,
The University of Pennsylvania, \\[-.1cm]
Philadelphia, PA 19104, USA} \\[.2cm]
3. \parbox[t]{6in}{Kavli Institute for Theoretical Physics,
University of California, \\[-.1cm]
Santa Barbara, CA 93106, USA}\\[.4cm]
\hbox{\hskip-1em
\email{sfranco@mit.edu,yanghe@physics.upenn.edu,herzog@kitp.ucsb.edu,walcher@kitp.ucsb.edu}}
}

\abstract{
We investigate the general features of renormalization group
flows near superconformal fixed points of four dimensional
${\cal N}=1$ gauge theories with gravity duals.  The gauge theories we
study arise as the world-volume theory on a set of D-branes
at a Calabi-Yau singularity where a del Pezzo surface
shrinks to zero size.  Based mainly on field theory
analysis, we find evidence that such flows are often chaotic
and contain exotic features such as duality walls. For a
gauge theory where the del Pezzo is the Hirzebruch zero
surface, the dependence of the duality wall height on the
couplings at some point in the cascade has a self-similar fractal 
structure.  For a gauge theory dual to
$\IP^2$ blown up at a point, we find periodic and quasi-periodic
behavior for the gauge theory couplings that does not
violate the $a$ conjecture.  Finally, we construct
supergravity duals for these del Pezzos 
that match our field theory beta functions.}
\begin{document}

\section{Introduction and Summary}

Understanding renormalization group flows out of conformal 
fixed points of supersymmetric gauge theories is of vital 
importance in fully grasping the AdS/CFT Correspondence beyond
super-conformal theories and
brings us closer to realistic gauge theories such as QCD. 
In particular, the ${\cal N}=1$ gauge theories 
arising from world-volume theories of D-brane
probes on Calabi-Yau singularities have been extensively
studied under this light. Dual to these theories are the
so-called non-spherical horizons of AdS
\cite{cones,MoPle}.

A prominent example, the conifold singularity,
was analysed by Klebanov and
Strassler (KS) in \cite{KS} where
the RG flow takes the form of a duality cascade. Here,
we have a theory with two gauge group factors and four
associated bi-fundamental fields. With the addition of
appropriate D5-branes, the theory is taken out of
conformality in the infra-red. Subsequently, the two gauge
couplings evolve according to non-trivial beta functions. 
Whenever one of the couplings becomes strong, we should perform 
Seiberg duality to migrate into a regime of weak coupling 
\cite{KS}. And so on do we proceed {\it ad infinitum}, 
generating an intertwining evolution for the
couplings. This is called the KS cascade.  The dual
supergravity (SuGRA) solution, happily aided by our full
cognizance of the metric on the conifold, can be studied in
detail and matches the field-theory behavior. 

One would imagine that a similar analysis, applied to more
general Calabi-Yau singularities than the conifold, could be
performed, {\it mutatis mutandis}.  Indeed, 
a full field theory treatment
can be undertaken using various techniques for constructing
the gauge theory for D-brane probes on wide classes of
singularities.
Behavior that differs dramatically from the
KS flow has been subsequently observed for, {\it exempli
gratia}, a class of non-spherical horizons which are $U(1)$
bundles over the del Pezzo surfaces
\cite{Hanany:2003xh,Franco:2003ja}. Using the
$a$-maximization procedures of \cite{IW,Intriligator:2003wr}
to determine anomalous dimensions and beta functions, the numerical 
studies of \cite{Franco:2003ja} have convinced us that, sensitive to
the type of geometry as well as initial conditions, the
quivers after a large number of Seiberg dualities may become
hyperbolic in the language of \cite{Fiol}. After this,
a finite energy scale is reached beyond which duality cannot 
proceed. This phenomenon has been dubbed a ``duality wall'' by
\cite{strassler}. 

The purpose of this paper was to
elucidate some aspects of flows, cascades, and 
walls for gauge theories arising from these more general 
geometries using both field theory and SuGRA techniques. 
To begin with, a more 
systematic, and where possible, analytic investigation of 
the duality wall phenomenon is clearly beckoning. For this 
purpose, we will use the exceptional collection techniques 
that become particularly conducive for the del Pezzo surfaces
\cite{unify}, especially for computing the beta functions
and Seiberg dualities \cite{H,HW}. We review these matters
synoptically in Section 2. In particular, we will formulate
the general RG cascade as motion and reflections in
certain {\bf simplices} in the space of gauge couplings.

Thus girt with the analytic form of the beta functions and 
Seiberg duality rules \cite{Franco:2003ja,H,HW}, we show 
in Section 3 the existence of the duality wall at finite 
energy. As an illustrative example, we focus on $F_0$, the 
zeroth Hirzebruch surface.  In the numerical studies 
of \cite{Franco:2003ja}, two types of cascading behavior 
were noted for $F_0$.  Depending on initial values of 
couplings, one type of cascade readily
caused the quiver to become hyperbolic and hence an
exponential growth of the ranks, whereby giving rise to a
wall. The other type, though seemingly asymptoting to a
wall, was not conclusive from the data. As an application of
our analytic methods, we show that duality walls indeed
exist for both types and give the position thereof as a
function of the initial couplings. These results represent the
first example in which the position of a duality wall along with all
the dual quivers in the cascade have been analytically determined. 
Thus, we consider it to be an interesting candidate to attempt
the construction of a SuGRA dual. Interestingly,
the duality wall height function is piece-wise linear
\cite{Hanany:2003xh,Franco:2003ja} and ``fractal.'' A highlight 
of this section will be the demonstration that a {\bf fractal}
behavior is indeed exhibited in such RG cascades. As we
zoom in on the wall-position curve, a self-similar pattern of
concave and convex cusps emerges. 

Inspired by this {\bf chaotic behavior}, we seek further in
our plethora of geometries for signatures of chaos. Moving
onto the next simplest horizon, namely that of the $dP_1$,
the first del Pezzo surface, we again study the analytic
evolution of the cascade in detail. Here, we find 
Poincar\'{e} cycles for trajectories of gauge coupling pairs. 
The shapes of these cycles depend on the initial
values of couplings. For some ranges, beautiful elliptical
orbits emerge. This type of behavior is 
reminiscent of the attractors and
Russian doll renormalization group flow discussed in 
\cite{chaos}.  This example
constitutes Section 4. 

Finally, in Section 5, we move on to the other side of the
AdS/CFT Correspondence and attempt to find SuGRA solutions.
We rely upon the methodologies of \cite{GP} to construct
solutions that are analogous to those of 
Klebanov and Tseytlin (KT) \cite{KT} for the
conifold. 
The fact that explicit metrics for cones over del 
Pezzo surfaces are not yet known is only a minor obstacle.
We are able to write down KT-like solutions, complete with
the warp factor, as an explicit function of the Cartan
matrices of the exceptional algebra associated with the del
Pezzo. 

These SuGRA solutions should be
dual to field theory cascades that are similar to the
original KS cascade.  Identifying the precise SuGRA phenomenon
which marks the duality wall remains an open and tantalizing
quest. 

We would like to stress the importance of possible
corrections to the R-charges of the matter fields, 
and hence to the 
anomalous dimensions and beta functions.
We will see that in order to be able to follow the
RG cascades accurately, we need to be able to assume
that the R-charges are corrected only at order
${\mathcal O}(M/N)^2$
where $M$ is the number of D5-branes, $N$ the number of D3-branes,
and $M/N$ a measure of how close we are to the conformal 
point $M=0$.  In the case of the conifold,
the gauge theory possessed a $\IZ_{2}$ symmetry which forced the
${\mathcal O}(M/N)$ corrections to vanish.
Our del Pezzo gauge theories generally lack such a symmetry.

We have two arguments to address these concerns.
First, for KS type cascades, our SuGRA solutions
 match the field theory 
beta functions precisely, severely constraining any
possible $M/N$ corrections to the R-charges.
For more complicated cascades involving duality walls,
we lack SuGRA solutions. 
Nevertheless, we shall push ahead, assuming that eventually
appropriate supergravity solutions will be found and that R-charge
corrections, even if ${\mathcal O}(M/N)$, will not change the
qualitative nature of our results.  The flows which we shall soon
present are so interesting that we
think it worthwhile to describe them in their current, though less
than fully understood state.
An analogy can be made to the Navier-Stokes equation.  Turbulence is
observed in fluids in many different situations but is very difficult
to model exactly.  Instead, people have developed simple models, such
as Feigenbaum's quadratic recursion relation, to understand certain
qualitative features, such as period doubling.  In some sense, the
flows we present here are in relation to the real RG flows as
Feigenbaum's analysis is to the real Navier-Stokes equation.

\section{A Simplicial View of RG Flow}

In preparation for our discussions on Renormalisation Group (RG) flow
in the gauge theory duals to del Pezzo horizons, we initiate our study
with an abstract and recollective
discussion of RG flows and duality cascades.

\subsection{The Klebanov-Strassler Cascade}

The Klebanov-Strassler (KS) flow \cite{KS}  provides our paradigm 
for an RG cascade.  In the KS flow, one starts
with an ${\mathcal N=1}$ $SU(N) \times SU(N+M)$ gauge theory
with bifundamental chiral superfields $A_i$ and $B_i$, $i=1,2$ and
a quartic superpotential. The couplings associated with the two gauge
groups we shall respectively call $g_1$ and $g_2$.
This quiver theory can be
geometrically realized as the world-volume theory of a stack of $N$
coincident D3-branes together with $M$ D5-branes 
probing a conifold singularity. The matter content and
superpotential are given as follows:
\beq
\label{conifold}
\ba{c}
\epsfxsize = 5cm
\epsfbox{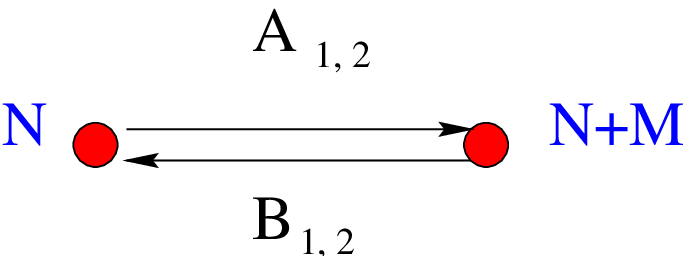}
\ea~~~~~~~~~
W = {\lambda \over 2} \epsilon^{ij} \epsilon^{kl} \Tr A_i B_k A_j B_l \ .
\eeq
where $\lambda$ is the superpotential coupling and the trace is 
taken over color indices.

For $M=0$ the gauge theory is conformal. Indeed, the $M$ D5-branes 
are added precisely to take us out of this conformal point, inducing a RG flow.

The one loop NSVZ beta function \cite{NSVZ} determines the running of the gauge 
couplings. For each gauge group we have
\beq
\beta_i = \frac{d(8\pi^2 /g_i^2)}{d \ln \mu}=
{3T(G)-\sum_i T(r_i)(1-2\gamma_i) \over 1-{g_i^2 \over 8 \pi^2} T(G) } 
\label{betadef}
\eeq
where $\mu$ is a ratio of energy scales 
and for an $SU(N_c)$ gauge group $T(G)=N_c$ and $T(fund)=1/2$.

Using $\gamma_i=\frac{3}{2} R_i-1$, we can express the beta functions $\beta_{i=1,2}$ for 
the two gauge couplings $g_{i=1,2}$ in terms of R-charges. 
As is done in \cite{KS}, we will work in the 
approximation that the denominator of \eref{betadef} is
neglected. Then, the  
beta functions become 
  
\begin{eqnarray*}
\beta_1 &=& 3\left[N + (R_A-1)(N+M) + (R_B-1)(N+M) \right] \ ,\\
\beta_2 &=& 3\left[(N+M) + (R_A-1)N + (R_B-1)N \right] \ .
\end{eqnarray*} 
 
At the conformal point, the R-charges of the bifundamentals can be calculated
from the geometry and are $R_A = R_B = 1/2$. They can also be simply determined
by using the symmetries of the quiver and requesting the vanishing of the beta
functions for the gauge and superpotential couplings. Generically, we
would expect  
the R-charges to suffer ${\mathcal O}(M/N)$ corrections for $M \neq
0$. Here however, 
there is a $\IZ_{2}$ symmetry $M \to -M$ for large $N$ that forces the
corrections  
to be of order at least ${\mathcal O}(M/N)^2$.  Thus,
\beq
\beta_1 = -3M, \qquad \beta_2 = 3M 
\eeq
up to ${\mathcal O}(M/N)$ corrections.

If we trust these one loop beta functions, then flowing into the IR,
we see that the coupling $g_2$ will eventually diverge because of the
positivity of $\beta_2$. According to Klebanov and Strassler, the appropriate 
remedy is a Seiberg duality. After the duality, the gauge
group becomes $SU(N) \times SU(N-M)$ but otherwise the
theory remains the same.  After this duality, the
beta functions change sign $\beta_1 = 3M$ and $\beta_2 = -3M$.
This process of Seiberg dualizing and flowing can be continued
for a long time in the large $N$ limit as shown in \fref{f:conicascade}.
The number of colors in the gauge groups becomes smaller and smaller.
Klebanov and Strassler \cite{KS} demonstrated
that when one of the gauge groups
becomes trivial, the gauge theory undergoes chiral symmetry 
breaking and confinement. 
The phenomenon is realized geometrically in the SuGRA 
dual by a deformation of the conifold in the IR.

\FIGURE[ht]{
  \epsfxsize = 3in
  \centerline{\epsfbox{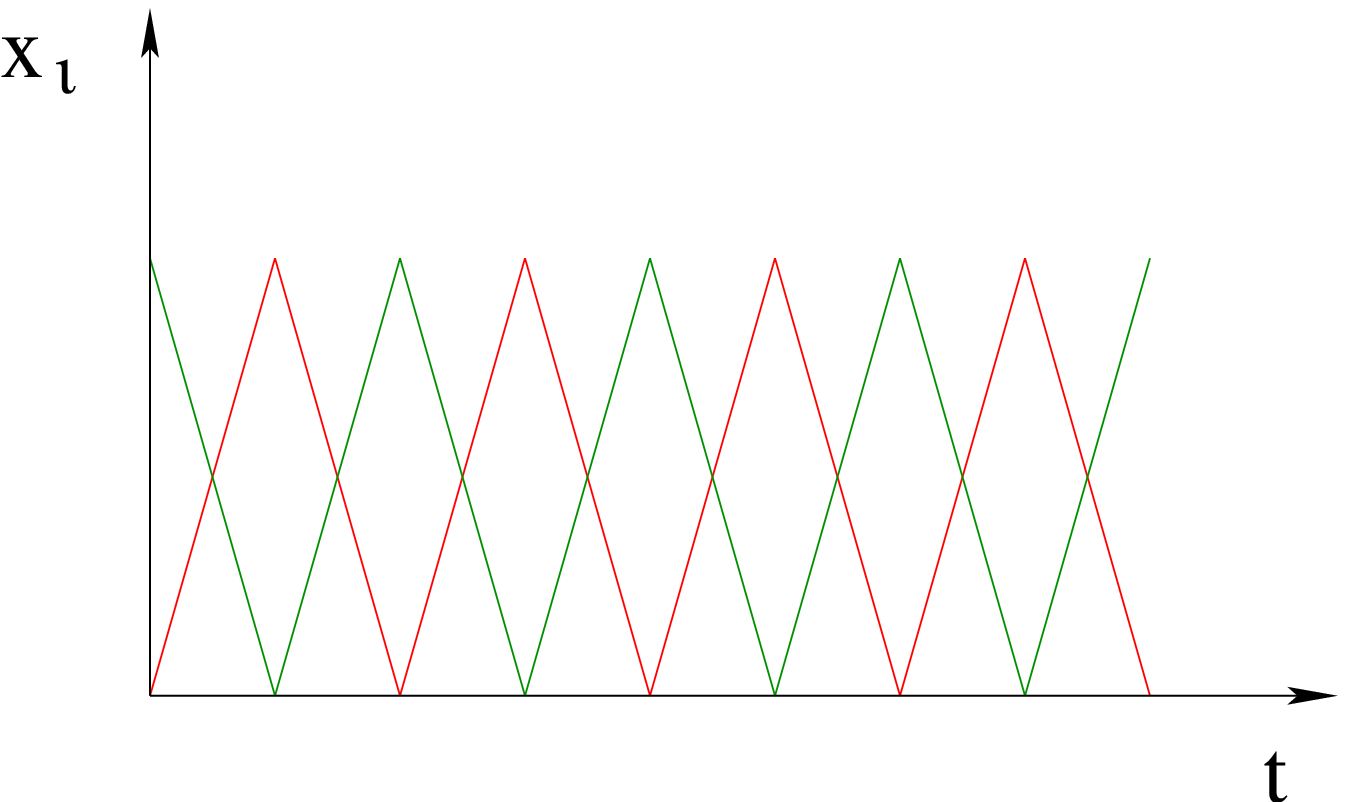}}
  \caption{{\sf The KS cascade for the conifold. The two inverse gauge
  couplings $x_{i=1,2} = \frac{1}{g^2_i}$ for the two nodes evolve in
  weave pattern against log-energy scale $t$ where Seiberg duality is
  applied whenever one of the $x_i$'s reaches zero.}}
  \label{f:conicascade}
}

Clearly there are some weaknesses in this purely gauge theoretic 
approach to the RG flow of a strongly coupled gauge theory.
Usually Seiberg duality is understood as an IR equivalence of
two gauge theories and is not performed in the limit $g_2 \to \infty$.
Can we really trust Seiberg duality here?  Also, we have dropped
the denominator of the full NSVZ beta function \eref{betadef}, which 
is presumably important. Nevertheless, the analysis is sound and
the strongest argument for the validity 
of these Seiberg dualities comes not from gauge theory but from
the dual supergravity theory \cite{KS}.  
There is a completely well-behaved supergravity solution, the KS
solution of the conifold, which models this RG flow. On the gravity side, 
there is a radial dependence of the 5-form flux which produces a logarithmic
running of the effective number of D3-branes in complete accordance with the
field theory cascade, giving credence to these Seiberg dualities.

\subsection{General RG Flows}
\label{general}

We shall henceforth focus on the four dimensional, ${\mathcal N}=1$ 
gauge theories engineered by placing D3-branes at the singularity of 
a Calabi-Yau threefold cone over a del Pezzo surface (cf.~e.~g.~
\cite{H,Greene,Feng:2000mi,Feng:2002zw,Franco:2002ae} 
for a comprehensive discussion). 
With some important caveats, these theories can be treated in a 
fashion similar to the discussion above for the conifold.

The field content of a del Pezzo gauge theory is described compactly
by a quiver.
For D-branes probing the $n$-th del Pezzo, the number of gauge
group factors in the quiver theory is equal to
\beq
k = n + 3 \ ,
\eeq
which is the Euler characteristic $\chi(dP_n)$. We reserve the
index $i=1,2,\ldots k$ for labeling the nodes of the quiver. 
We denote the adjacency matrix of the quiver by $f_{ij}$. In
other words, $f_{ij}$ is the number of arrows in the quiver from
node $i$ to node $j$. We point out that by definition, the
$f_{ij}$ are all non-negative.

Thus given a quiver, we need to specify the ranks of the gauge
groups in order to define a gauge theory. We will denote the
rank of the gauge group on the $i$-th node by $d^i$, and the
dimension vector by $d=(d^i)_{i=1,\ldots,k}$. As on the conifold, 
the ranks $d^i$ are related to the number of branes that realize 
the specific gauge theory in string theory. When probing the
del Pezzos, we will reserve $N$ to denote the number of
regular D3-branes, and $M^I$ to denote the number of D5-branes. 
The D3-brane corresponds to a unique dimension vector which
we will denote by $r=(r^i)_{i=1,\ldots k}$. In distinction to
the conifold and its ADE generalizations, the possible
D5-branes are constrained by chiral anomaly cancellation,
and we will parametrize their dimension vectors by $s_I=(s^i_I)$
with $I=1,2\ldots,n$.

Summarizing, a D-brane configuration with $N$ regular D3-branes
and $M^I$ D5-branes of type $I$ corresponds to the gauge
group $\prod\limits_{i=1}^k SU(d^i)$ with
\beq
d^i  = r^i N + s^i_I M^I
\eeq
and $f_{ij}$ chiral fields $X_{ij}$ in the $SU(d^i)\times 
SU(\overline{d^j})$ bi-fundamental representation.

As shown in \cite{H,HW}, the beta functions of the gauge theory can
be computed effectively from geometry by taking advantage of the 
exceptional collection language \cite{unify, H, HW, soliton}.
An exceptional collection $\cale=(E_1,E_2,\ldots,E_k)$ is an ordered 
collection of sheaves, specifying the D-brane associated with
each node. The intersections of the sheaves give rise to massless 
strings which in turn correspond to bifundamental fields in the 
gauge theory. $\cale$ can roughly be thought of as a basis of branes.

An important feature of exceptional collections for us will be
the ordering.  The ordering of a collection induces an ordering of
the nodes of the quiver. 
In order to use the exceptional collection technology to compute the
beta functions, we must keep track of the ordering.

If a given quiver satisfies the well split condition of \cite{H}, the
order of the quiver changes in a simple way under Seiberg duality.  To
understand the well split condition, we first need to refine 
our understanding of the quiver ordering.  It was shown in \cite{H}
that the ordering of the 
quiver is only determined up to cyclic permutations.  If $123\ldots n$
is a good ordering, 
then so is $23 \ldots n1$.  If a quiver is well split, then we can
find a cyclic permutation 
such that for any node $j$, all the outgoing arrows from $j$ go to
nodes $i<j$ and all the 
in-going arrows into $j$ come from nodes $i>j$.
After a Seiberg duality on node $j$, $j$ would become the last node in
the quiver. 

An unproven conjecture of \cite{H} is that the Seiberg dual of a well
split quiver is again well split. The conjecture was proven for four 
node quivers in \cite{H} and no counter-examples are known to 
the authors. An appropriate understanding of ill split quivers
is still lacking. For example, the correct determination of
R-charges for them is still open \cite{H}. Indeed, the fractional 
Seiberg dualities encountered in \cite{soliton} may be problematic 
precisely for this reason. As our examples in the subsequent sections 
involve only Seiberg dualities of well split, 
four node quivers, we can be confident 
in our calculations.

In light of the exceptional collection language, we shall also make
use of the matrix $S$ which is an upper triangular matrix with
ones along the diagonal and related to $f_{ij}$ by
\beq
\label{sa}
S_{ij} = 
\left\{
\ba{lll}
f_{ij} - f_{ji} \ , & & i < j \ ; \\
1 \ , & & i = j \ ; \\
0 \ , & & i > j \ .
\ea \right.\,,
\eeq
where we have assumed an ordering. The components $S_{ij}$, $i\neq j$, 
are still the number of arrows from node $i$ to node $j$, except that
now a negative entry corresponds to reversing the arrow direction.
We will find it convenient to use a matrix $\cali$ which is
simply the antisymmetrized version of $S$ (or $f$).
\beq
\cali = S - S^t = f-f^t
\eeq
Using this, chiral anomaly cancellation can be concisely expressed 
as the condition that the dimension vector $d$ be in the kernel
of $\cali$. In other words, $r$ and the $s_I$ form a basis of 
$\ker \cali$.

\subsubsection{Beta Functions and Flows}

Methods exist in the literature for the determination of the R-charges
as well as the beta function. Evaluating (\ref{betadef}) with the
quiver notation introduced above, and denoting by $R_{ij}$ the
R-charge of the bifundamental $X_{ij}$, one obtains for the beta
function of the $i$-th node (cf. Eq (5.7) of \cite{Franco:2003ja})
\beq
\label{beta1}
\frac{d x_i}{d \ln \mu} = 
\beta_{i}
= \left(3 d^i + \frac32
        \sum_{j=1}^k \left(f_{ij} (R_{ij}-1) + f_{ji}(R_{ji}-1)
\right) d^j \right)\ . 
\eeq
where $x_i$ is related to the $i$-th gauge coupling via 
$x_i\equiv 8\pi^2/g_i^2$.

One very insightful approach for the determination of the R-charge
is the procedure of maximization of the central charge $a$ in the CFT
as advocated in \cite{IW,Intriligator:2003wr}. We shall however adhere
to the procedure of \cite{H,HW}, which gives the R-charges at the
conformal point. Transcribing Eq.\ 49 from \cite{HW} to present 
notations, the R-charge of the bi-fundamental $X_{ij}$ is given by
\beq
\label{R}
R(X_{ij}) = 1+\left(
\frac{2}{(9-n)r^i r^j} (S^{-1}_{ij} + S^{-1}_{ji}) - 1 
\right) \sign(i-j) \ .
\eeq
It was shown in \cite{HW} that plugging (\ref{R}) into
(\ref{beta1}), and going to the conformal point $d^i=r^i$, one finds
$\beta_i=0$, as expected.

The flow is induced when we leave the conformal fixed
point by adding D5-branes. 
As in \cite{KS}, we will work in the regime $M^I\ll N$.
We will assume the R-charges do not receive 
corrections of 
${\cal O} (M^I/N)$.  
This assumption is 
supported by the supergravity solutions
we write down in section 5, which severely
constrain the nature of such corrections for
KS type cascades. 
For more general cascades with duality walls,
we believe that we can still trust the qualitative nature
of our results.
Ignoring the corrections, the
non-conformal beta functions can readily be obtained by
substituting \eref{R} into \eref{beta1} for general ranks $d^i$.
We obtain, to order $M^I/N$,
\beq
\beta_i = 3s^i_I M^I + \frac{3}{2} \sum_{j} 
\Rt_{ij} s^j_I M^I \,,
\label{beta}
\ee
where we have introduced the symmetric matrix
\beq
\label{Rt}
\Rt_{ij} = f_{ij} (R_{ij}-1) + f_{ji} (R_{ji}-1) \ .
\eeq 

We will now evolve the inverse gauge couplings $x_i=
8\pi^2/g_i^2$ with the beta functions \eref{beta}.
Since the one-loop beta functions are constant, 
the evolution proceeds in step-wise linear fashion, much
like the KS cascade; we have
\beq
\frac{8\pi^2}{g_i^2(t + \Delta t)} - \frac{8\pi^2}{g_i^2(t)} = \beta_i \Delta
t
\eeq
during the step $\Delta t$ in energy scale ($t=\ln \mu$), 
before one has to perform
Seiberg duality on the node whose coupling reaches zero first.

An important constraint can be placed on this evolution.
Even though now these beta functions do not vanish identically, it is
still the case that 
\beq\label{betasum}
\sum_i^k \beta_i r^i = 0 \ .
\eeq
The reason is that this sum can be reorganized into a sum over
each of the beta functions at the conformal point, and at 
the conformal point, each of these beta functions vanishes
individually.
It follows from \eref{beta} and \eref{betasum} therefore, that, 
\be
\sum_i^k \frac{r^i}{g_i^2} = \mbox{constant} \ 
\label{simprel}
\ee
throughout the course of the cascade; on this constraint we shall
expound next.

\subsubsection{Simplices in the Space of Couplings}
\label{s:simp}

The space of possible gauge coupling constants $x_i \equiv 1/g_i^2$ 
for a quiver with $k$ gauge groups is a cone $(\IR_+)^k$.
The relation (\ref{simprel}) cuts out a {\bf simplex} in this 
cone.  The beta functions (\ref{beta}) establish the direction
of the renormalization group flow inside this simplex.
For the KS conifold flow, having two gauge couplings, the cone is the 
quadrant in $\IR^2$ parametrized by $1/g_1^2 = x>0$ and $1/g_2^2 = y>0$.  
The simplex is the line segment $x+y=\mbox{const}$ inside this cone. 
The beta functions tell us to move up and down this line segment until
one or the other coupling constant diverges.

In more general cases, under the renormalization group flow, we will eventually
reach a face of the simplex where one of the couplings
diverges.  At this point, the insight gained from the KS flow tells
us we should Seiberg dualize the corresponding gauge group.  After the duality, we 
find ourselves typically in a new gauge theory.  The new gauge theory
has some new associated simplex and renormalization group
flow direction given by some different set of beta functions.  The KS flow is
very special in that the Seiberg dual theory is identical to the original one
up to the total number of D3-branes $N$.  

One imagines in general some huge collection of simplices glued together
along their faces.  In any given simplex, the renormalization group 
trajectory is a straight line.  At the faces, the trajectory
``refracts''.  One recomputes the beta functions to find the new direction 
for the RG flow. In \fref{f:conicascade} for example, we have the evolution of
the couplings reflecting off the $t$-axis (corresponding to either $1/g_1^2$ 
or $1/g_2^2$ equal to zero), whereby giving the weave pattern.
Note that such RG flows are generically quite sensitive to initial
conditions. Slightly altering the initial couplings may alter the
trajectory such that a different face of a simplex is reached.  A
different face corresponds to a Seiberg duality on a different node
which will generically completely alter the rest of the flow. Such
a sensitivity was noticed in \cite{Hanany:2003xh, Franco:2003ja}.

For four node quivers, the simplices are tetrahedra and the RG flow
can be visualized.  There is only one vector $s$, with
components $s_i$, corresponding to
only one D5-brane.  Thus, the direction of the RG flow inside
any given tetrahedron is, up to sign, independent of $M$.
Moreover, one can show that after a duality on node $i$, $\beta_i \to
-\beta_i$ (see the appendix for details).

Thus prepared, we can embark upon a detailed study of the RG flows and
duality cascades for various concrete examples. Some of them will 
exhibit a KS type behavior, meaning that the cascade will periodically
return to the same quiver up to a change in the number of D3-branes, showing
no accumulation of dualization scales in the UV. Others will be markedly 
different, exhibiting duality walls. In particular, we shall describe an 
assortment of interesting flows for D-branes probing cones over the del 
Pezzo surfaces, where we will be able, in addition to numerics, to
gain some quantitative analytic understanding.

\section{Duality Walls for $F_0$}

We begin with D-brane probe theories on the complex cone over $F_0$, the 
zeroth Hirzebruch surface. The addition of D5-branes takes us out 
of conformality, whereby inducing a RG flow. Detailed numerical study
was undertaken in \cite{Franco:2003ja}. All Seiberg dual theories for this
geometry can be arranged into a web which encodes all possible duality 
cascades. This web takes the form of a flower and has been affectionately 
called the {\it Flos Hirzebruchiensis} (cf.~Fig 7 cit.~Ibid.). The purpose of this section is to derive analytical results for the existence of duality 
walls and their location.  We also explain the
{\bf fractal structure} of the duality
wall curve as a function of the initial couplings. 
%
\subsection{Type A and Type B Cascades}

Before proceeding with the analytical derivations, let us make a brief
summary of the findings in \cite{Franco:2003ja}, 
where two classes of RG trajectories were identified. 
In one gauge theory realization,
$F_0$ exhibits a Klebanov-Strassler type flow that alternates between
two quivers with  
constant intervals in $t = \log \mu$ (for energy scale $\mu$)
between successive dualizations. This type
of flow is an immediate generalization of the  
conifold cascade. The quivers of and the beta functions inter-connecting
between the two theories are shown in \eref{f:F0-1}.
\FIGURE[ht]{
  \epsfxsize = 6in
  \centerline{\epsfbox{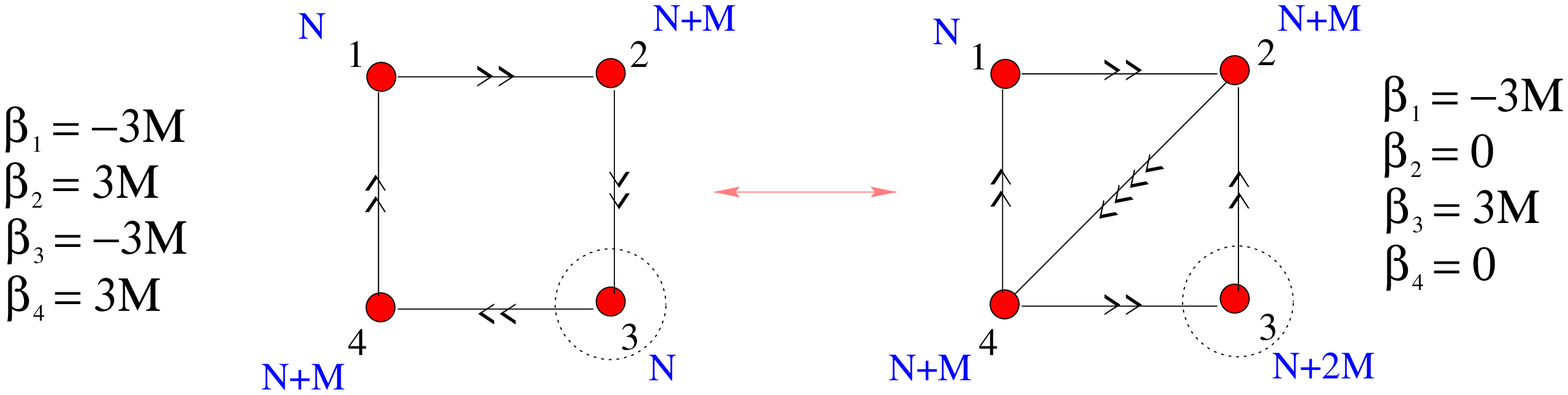}}
  \caption{{\sf The first class of duality cascades for $F_0$. This is
  an immediate generalization of the KS conifold case and we
  alternate between the two theories upon dualizing node 3 of each and
  evolve according to the beta functions shown.}}
  \label{f:F0-1}}

The second class of flows commences with the quiver in \fref{f:F0-2}, which
is another theory in the duality flower for $F_0$.
In this case, there is a decrease in the $t$ interval between
consecutive dualizations towards the UV, leading to the possibility of
a so-called ``duality wall'' past which no more dualization is possible
and we have an accumulation point at finite energy.
Considering initial couplings of the four gauge group factors 
of the form $(1,x_2,x_3,0)$, two
qualitatively different behaviors were observed. 
\FIGURE[ht]{
  \epsfxsize = 3.6in
  \centerline{\epsfbox{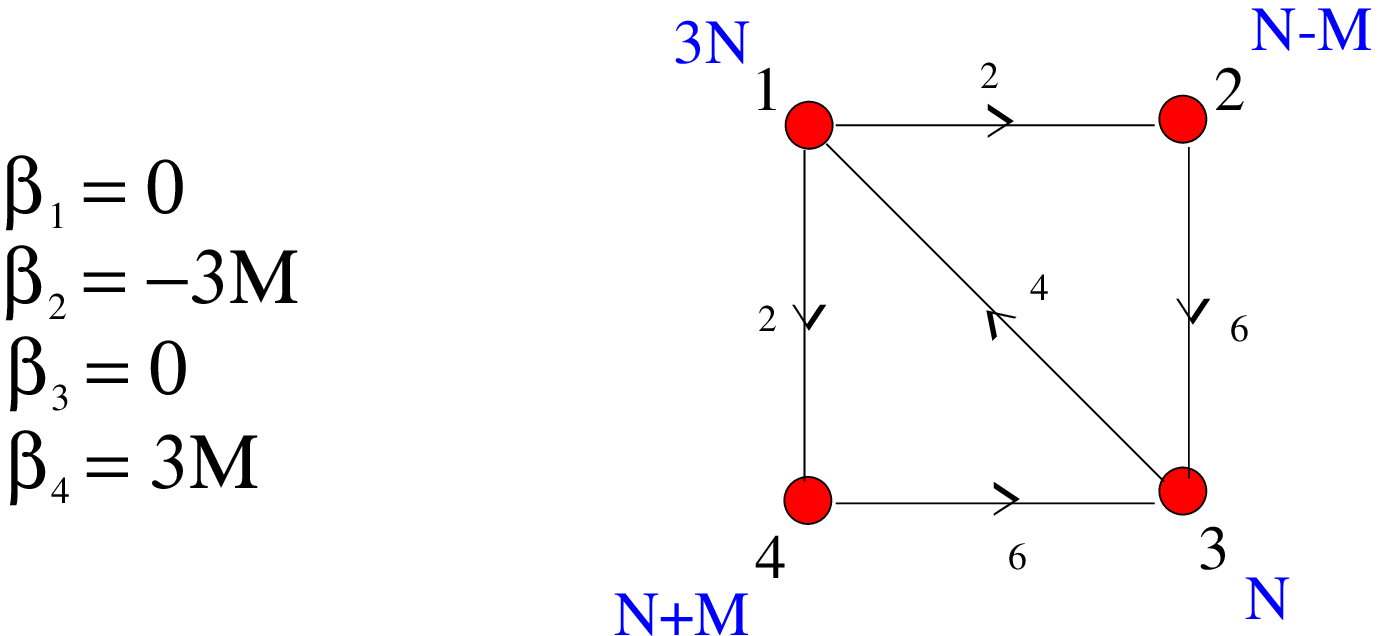}}
  \caption{{\sf The second class of theories for $F_0$. Starting from
  this quiver and following the duality cascade give markedly
  different behavior from the KS case. It was seen in this case that 
  the increment in energy scale decreases at each step and a ``duality wall'' 
  may be reached \cite{Franco:2003ja}.}}
  \label{f:F0-2}}
\begin{enumerate}
\item
In theories with $x_3 > 0.9$, the cascade
corresponds to an infinite set of  
alternate dualizations of nodes $1$ and $2$. The distance between
dualizations is monotonically decreasing, as was shown in Figures 12
and 13 of \cite{Franco:2003ja}. However, no  
conclusive evidence of convergence to a duality wall was found 
therein. 
We will call such a cascade an {\bf A type cascade} and  will show
shortly that in
this case a duality wall is indeed approached smoothly. 
\item
On the other hand, for $x_3 < 0.9$,  
the third gauge group is dualized at a finite scale. When this
happens, all the intersection
numbers in the quiver become 
larger than $2$, leading to an explosive growth of the ranks of the
gauge groups and the number of bifundamental chiral fields, 
and generating an immediate accumulation of the dualization
scales. This discontinuous behavior makes duality walls 
evident even in numerical simulations for these flows. We will refer to
these flows as {\bf B type cascades}.
\end{enumerate}

\subsection{Duality Walls in Type A Cascade}
\label{section_quivers}

Having elucidated the rudiments of the cascading behavior of the
$F_0$ theories, let us explore whether there are indeed duality walls
for A type cascades, which we recall to be the type for which
numerical evidence is not conclusive. We shall proceed analytically.
In order to do so, let us first construct the quiver at an arbitrary
step $k$. We can regard Seiberg duality as a matrix
transformation on the rank vector and the adjacency matrix as was done
for example in Sec.~8.1 of \cite{Franco:2003ja}.
An elegant way to derive the quiver at a generic position in
the cascade is by realizing Seiberg duality transformations 
as mutations in an exceptional collection (equivalently, by
Picard-Lefschetz monodromy transformations on the 3-cycles in 
the manifold mirror to the original Calabi-Yau).
We will use this language as was done in \cite{H,HW}.

Taking the
exceptional collection to be $(a,b,3,4)$, the alternate dualizations
of nodes 1 and 2 corresponds in this language to the repeated left
mutation of $a$ 
with respect to $b$. For even $k$ $(a,b)=(1,2)$, while for odd $k$
$(a,b)=(2,1)$.  \fref{f:F0-2} corresponds to $k=1$ where 
the exceptional collection ordering is (2,1,3,4).  This quiver is well split.

%
\subsubsection{Quivers at Step $k$}
\label{section_quivers2}

Under Seiberg duality, the rank of the relevant gauge
group changes from $N_c$ to $N_f - N_c$. Type A cascades
correspond to always dualizing node $a$. By explicitly
constructing these RG trajectories, we will check that this
assumption is indeed consistent. The exceptional
collection tells us that after the duality, nodes $a$
and $b$ will switch places.  Thus
\be
\begin{array}{l}
N_a(k+1) = N_b(k) \ , \\
N_b(k+1) = 2N_b(k) - N_a(k) \ . 
\end{array}
\ee
It is immediate to prove that after $k$ iterations, the ranks of the
$SU(N_i)$ gauge groups are given by
\beq
\begin{array}{l}
N_a=(2k-1)N+(k-2)M \ ,\\
N_b=(2k+1)N+(k-1) M \ ,\\
N_3=N \ ,\\
N_4=N+M \ .
\end{array}
\label{ranks_odd}
\eeq
The number of bifundamental fields between each pair of nodes follow
from applying the usual rules for Seiberg duality of a quiver theory.
In particular, we combine the bifundamentals $X_{a4}$, $X_{ba}$,
and $X_{a3}$ into mesonic operators $M_{b4}= X_{ba}X_{a4}$ and
$M_{b3} = X_{ba} X_{a3}$.  We introduce new bifundamentals
$X_{4a}'$, $X_{ab}'$, and $X_{3a}'$ with dual quantum numbers along
with the extra term $M_{b4} X_{4a}' X_{ab}' + M_{b3} X_{3a}' X_{ab}'$
to the superpotential. 
We then use the superpotential to integrate out the massive fields, which
appear in the quiver as bidirectional arrows between the pairs
of nodes $(3,b)$ and $(4,b)$.
The resulting incidence matrix for the quiver will change such that
\beq
\begin{array}{lclcl}
f_{ba}(k+1)=f_{ba}(k)      & \ \ \ \ \ & f_{3b}(k+1)= f_{a3}(k)& \ \ \
\ \ & f_{43}(k+1)=f_{43}(k) \\ 
f_{a4}(k+1)=-f_{4b}(k) + 2 f_{a4}(k) & \ \ \ \ \ &
f_{4b}(k+1)=f_{a4}(k) & \ \ \ \ \ &  
f_{a3}(k+1)=-f_{3b}(k)+2f_{a3}(k)
\end{array}
\eeq
which can be simplified to yield
\beq
\label{F0-adj}
\begin{array}{lclcl}
f_{ba}(k)=2      & \ \ \ \ \ & f_{3b}(k)=2(k+1) & \ \ \ \ \ & f_{43}(k)=6 \\
f_{a4}(k)=2(k-1) & \ \ \ \ \ & f_{4b}(k)=2(k-2) & \ \ \ \ \ & f_{a3}(k)=2(k+2) \ .
\end{array}
\eeq
This information can be summarized in the quiver diagram in
\fref{quiver_F0}. 
\FIGURE[ht]{
  \epsfxsize = 6cm
  \centerline{\epsfbox{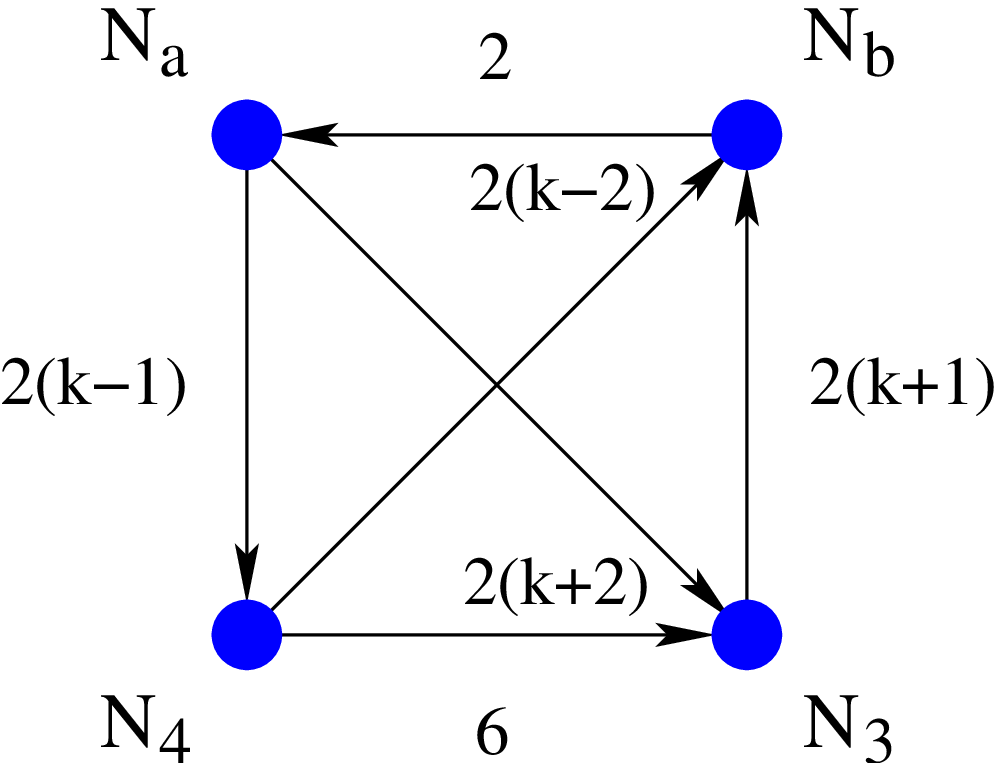}}
  \caption{{\sf Quiver diagram at step $k$ of a type A cascade for $F_0$.}}
  \label{quiver_F0}}

With the adjacency matrix \eref{F0-adj} and the non-conformal ranks
\eref{ranks_odd}, we can readily compute the
beta functions from \eref{beta}, to arrive at
\beq
\ba{lll}
\begin{array}{ll}
\beta_a=-{9 (k+1) k M \over (4k+2)}& <0 \\
\beta_b={9 (k-1) k M \over (4k-2)} & \geq 0\\
\beta_3={3 (7k^2-3k-4) M \over (2-8k^2)} & <0\\
\beta_4={3 (7k^2+3k-4) M \over (-2+8k^2)} & >0 \ ,
\end{array}
& \qquad &
\ba{l}
k = 1,2,3, \ldots \\
(a,b) = (2,1); \qquad k~\mbox{odd} \\
(a,b) = (1,2); \qquad k~\mbox{even}
\ea
\ea
\label{beta-F0}
\eeq

%
\subsubsection{The RG Flow}
\label{sectionRGflow}
Using the results in Section \ref{section_quivers2}, we proceed to
study the evolution of the dualization scales starting with the initial
couplings $(1,x_2(0),x_3(0),0)$.
Let us consider the first step in the cascade. 
We are in a type A cascade, so $x_3(0) > 0.9$.
The beta functions are, from \eref{beta-F0},
\beq
\beta_1(1) = 0, \ \ \beta_2(1) = -3M, \ \
\beta_3(1) = 0, \ \ \beta_4(1) = 3M \ .
\eeq
We see that only node 2 has a negative beta function at the first step
and so its associated coupling will reach zero first, i.e.,
the first step ends with the dualization of node 2. 
The subsequent increment $\Delta(1)$
in the energy scale $t = \log \mu$ before the dualization is performed
is equal to
\beq
\Delta(1)={x_2(0) \over |\beta_2(1)|} \ .
\eeq
Applying
\beq
x_i(k+1) = x_i(k) + \beta_i(k+1) \Delta(k+1), 
\quad t(k+1) = t(k) + \Delta(k) \ ,
\eeq
we have at the end of this step
\beq
x_1(1)=1, \  x_2(1)=0, \ x_3(1)= x_3(0), \
x_4(1)=\frac{3 M x_2(0)}{|\beta_2(1)|} \ .
\eeq
So, as far as nodes 2 and 3 are concerned,
the initial value $x_2(0)$ only affects the length of the first
step, beyond which any information about it is erased.
In order to look for the initial
couplings that lead to a type A flow,  recall that
we have to determine the possible  initial values $x_3(0)$ such that
$x_3(k)$ remains greater than zero as  
$k \rightarrow \infty$ so that the third node never becomes dualized. 
Since $\beta_3(1)=0$, this is completely
independent of $\Delta(1)$  and hence independent of $x_2(0)$.

That said, let us look at the cascade at the next step. 
The beta functions \eref{beta-F0} now give
\beq
\beta_1(2) = -\frac{27}{5}M, \ \ \beta_2(2) = 3M, \ \
\beta_3(2) = -\frac95M, \ \ \beta_4(2) = 3M \ .
\eeq
Since we are interested in type A cascades, we assume that the initial value 
$x_3(0)$ is such that this node is never dualized. Thus, the next node
to undergo Seiberg duality is the other one with a negative beta function, 
namely node 1. Recalling that $x_1(1)=1$, 
the consequent step in the energy scale $\Delta(2)$ is thus
\beq
\Delta(2)={x_1(1) \over |\beta_1(2)|} = \frac{1}{|\beta_1(2)|} \ ,
\eeq
and $x_1(2)=0$ while $x_2(2) = \beta_2(2) \Delta(2)$.
Proceeding similarly, the next step gives 
\beq
\Delta(3)={\beta_2(2) \over \beta_1(2)} {1 \over \beta_2(3)} \ .
\eeq
We see that in general, at the $k^{th}$ step, the interval $\Delta(k)$
is given by
\beq
\label{delta1}
\Delta(k)=\left[\prod_{i=2}^k  {\beta_b(i) \over |\beta_a(i)|}\right]
\frac{1}{\beta_b(k)} \ , \qquad
\ba{l}
(a,b) = (2,1), \quad k~\mbox{odd}; \\
(a,b) = (1,2), \quad k~\mbox{even} \ ,
\ea
\eeq
for $k \ge 2$.
This, using \eref{beta-F0}, can be written as a telescoping product
\beq
M \Delta(k)=\left[ \prod_{i=2}^k {(i-1) \over (i+1)} {(2i+1) \over
(2i-1)} \right] {(4k-2) \over 9 (k-1) k} \ .
\eeq
Simplifying this expression we arrive at
\beq
M \Delta(k)={2 (2k+1) (4k-2) \over 27 k^2 (k^2-1)}
\label{delta}
\eeq
for $k \geq 2$. 
The total variation of the third coupling $x_3$, after $k$ steps, 
is given by
\beq
x_3(k)-x_3(0)=\sum_{i=2}^k \Delta(i) \beta_3(i) \ .
\label{x3_step_k}
\eeq
As discussed, the boundary between type A and B cascades corresponds
to initial conditions such that $x_3(k) \rightarrow 0$ for
$k \rightarrow \infty$, i.e., the initial conditions
that separate the regime in which node 3 gets dualized at some finite
$k$ from the one in which it never undergoes a Seiberg duality. Then,

\beq
x_3(0) - x_{3}(\infty) = 
{2 \over 9} \sum_{i=2}^\infty {(7 i+4) \over i^2 (i+1)}
= {4 \over 27} \pi^2-{5 \over 9} \ .
\label{x3b1}
\eeq
We see that this sum is approximately equal to $0.906608$, in agreement 
with the numerical evidence, which located the transition at $x_3(0) \sim 0.9$.
We will henceforth call this coupling $x_3(0)$ such $x_{3}(\infty)=0$, $x_{3b}$, 
because it is a boundary value between type A and type B cascades.

\subsubsection{Duality Walls in Type A Cascades}

The computations in the previous section enable us to address one of
the questions left open in \cite{Franco:2003ja}, namely 
whether duality walls exist in this case. Our flow, from \eref{x3b1}, 
corresponds to an infinite cascade that only involves nodes 1 and 2.
Let us sum up all the steps $\Delta(k)$ in the energy scale {\it ad
infinitum}; this is equal to
\beq\label{sumdel}
\sum_{k=1}^\infty \Delta(k)=\Delta(1)+\sum_{k=2}^\infty \Delta(k) \ .
\eeq
Using $\Delta(1)=x_2(0)/|\beta_2(1)|=x_2(0)/3M$ and (\ref{delta}), we see
that this sum can actually be performed, giving us a finite
answer. This means that there is indeed a duality wall for our type A
cascades, whose value is equal to
\beq
t_{wall} = {1 \over 3M} \left( x_2(0)+{2 \pi^2 \over 27}+{5 \over 9}
\label{wall_type_A}
\right) \ .
\eeq
We would like to emphasize that, although derived in the approximation
of vanishing ${\mathcal O}(M/N)$ corrections to the R-charges, 
(\ref{wall_type_A})
is the first analytical result for a duality wall. Given the detailed
understanding we have of every step of the cascade on the gauge theory 
side, this example stands as a natural candidate in which to try to 
look for a realization of this phenomenon in a SuGRA dual.


\subsection{Fractal Structure of the Duality Wall Curve}

Having analytically ascertained the existence and precise position of the duality wall for type A cascades, 
and the boundary value $x_{3b}(0)$ of the inverse squared coupling at which the cascades become type B, we 
now move on to address another fascinating question, hints of which were raised in \cite{Franco:2003ja,HW}, 
viz., the dependence of the position of the wall upon the initial couplings. We will see that, in type B 
cascades, such dependence takes the form of a {\em self-similar} curve.

Let us focus on the one dimensional subset of the possible initial conditions given by couplings of the form 
$(1,1,x_3(0),0)$ (more general initial conditions can be studied in a similar fashion). \fref{wall_F0} is a 
plot of the position of the duality wall as a function of $x_3(0)$. Initial values $x_3(0)>x_{3b}$ correspond to 
type A cascades. Node 3 is not dualized in this case and thus the position of the wall is independent of 
$x_3(0)$ in this range, as determined by \eref{wall_type_A}. From now on, we will focus on the $x_3(0)<x_{3b}$ type B region. 
The curve exhibits in this region an apparent piecewise linear structure as was noticed in \cite{Franco:2003ja}.

\FIGURE[ht]{
  \epsfxsize = 9cm
  \centerline{\epsfbox{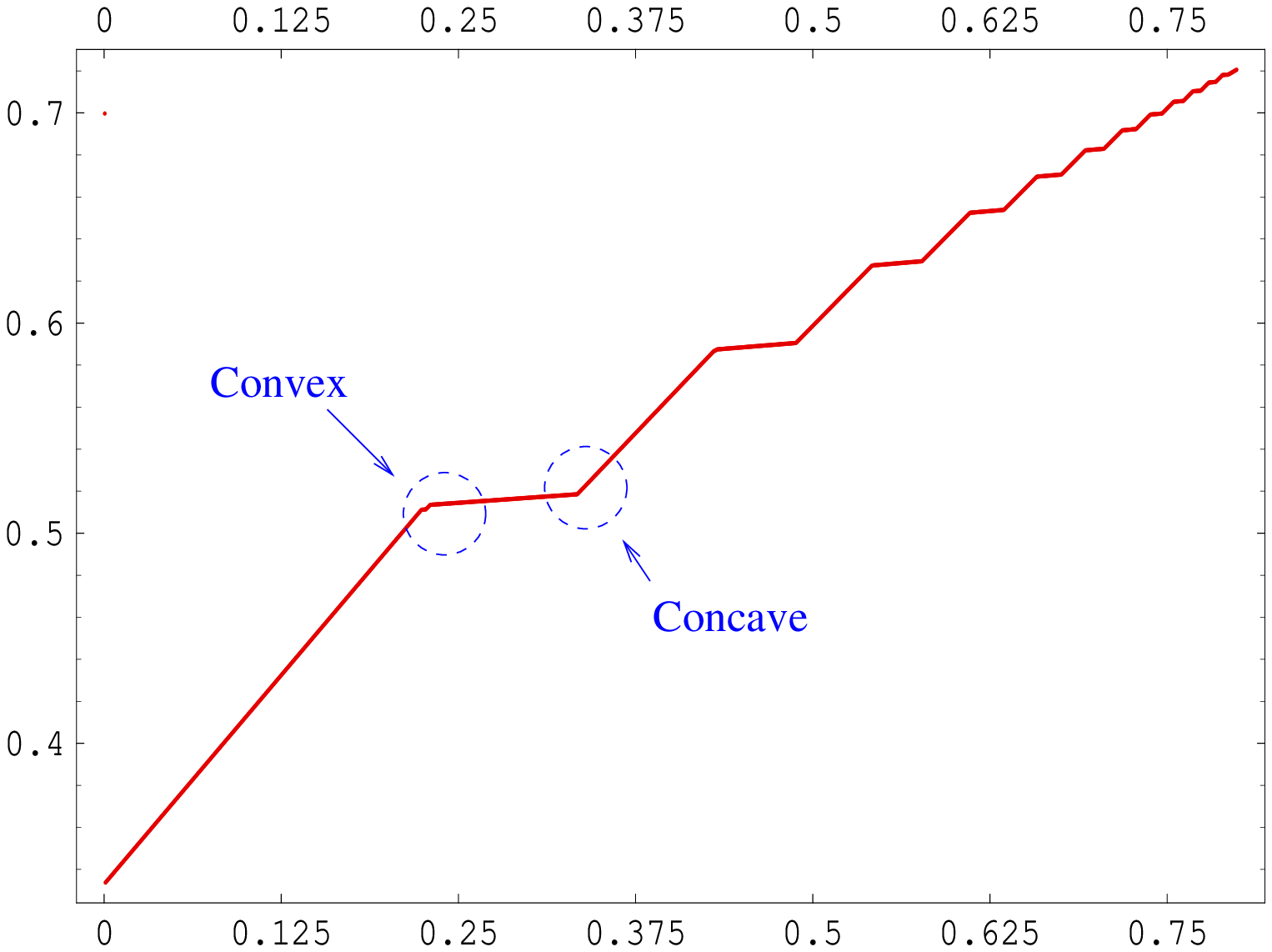}}
  \caption{{\sf Position of the duality wall for $F_0$ as a function
  of $x_3(0)$ for initial conditions of the  
                form $(1,1,x_3(0),0)$. A piecewise linear structure is
  seen for the type B cascade region,  
                i.e., $x_3(0)<x_{3b} \sim 0.9$.}}
  \label{wall_F0}}

In order to appreciate the piecewise structure more clearly, it is useful to consider the derivative of the curve. We 
show in \fref{dwall_F0} a numerical differentiation of \fref{wall_F0}. This apparent linearity is in fact approximate,
and an intricate structure is revealed when we look at the curve in more detail. While exploring the origin of the 
different features of the curve, we will discover that a {\em self-similar fractal structure} emerges.

\FIGURE[ht]{
  \epsfxsize = 9cm
  \centerline{\epsfbox{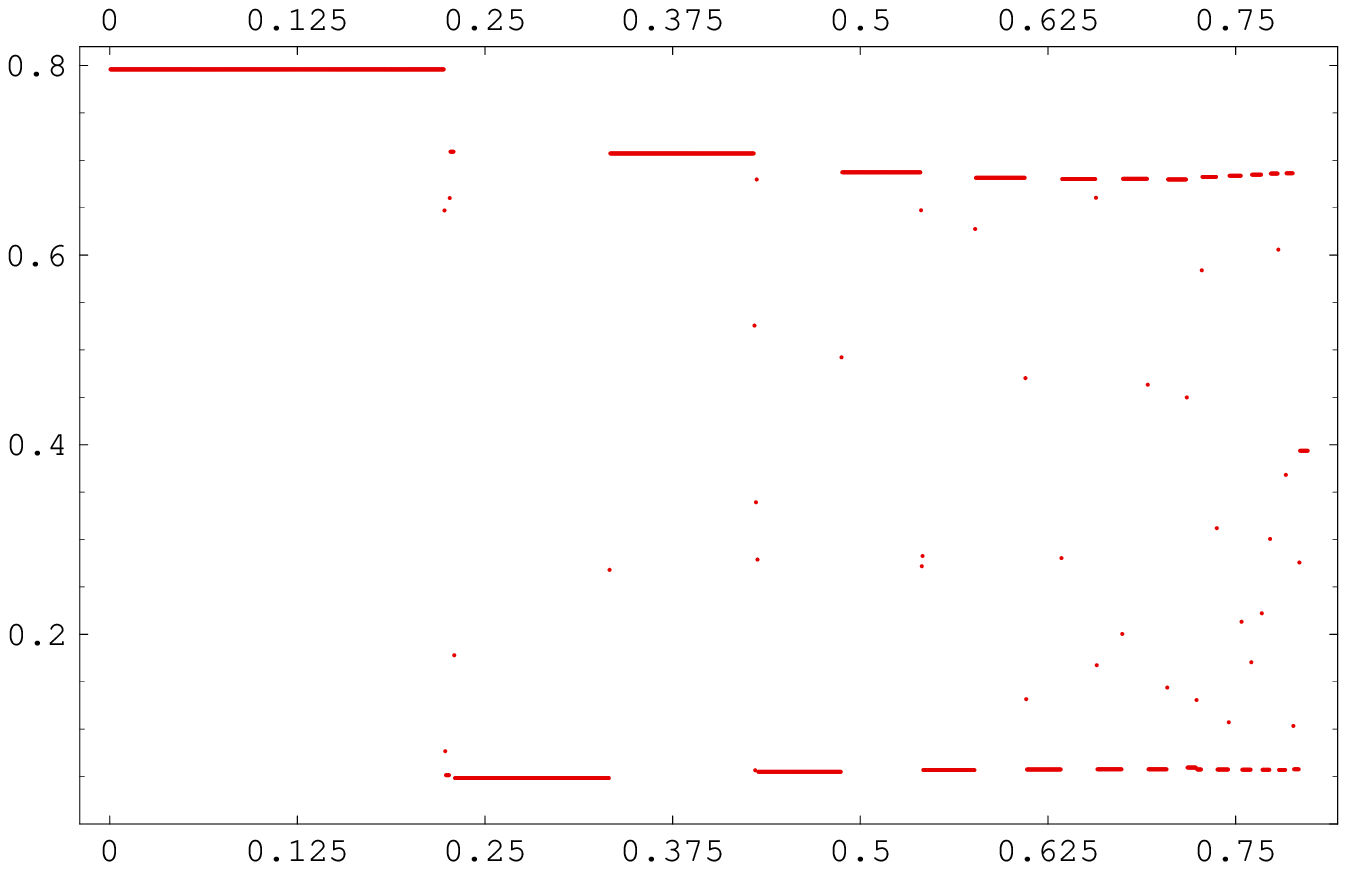}}
  \caption{{\sf Derivative of the position of the duality wall for
  $F_0$ as a function of $x_3(0)$ for initial  
                conditions of the form $(1,1,x_3(0),0)$. The
  appearance of the constant segments evidences 
                further the piecewise linear behavior of position of
  the wall with respect to $x_3(0)$.}} 
\label{dwall_F0}}

The most prominent features in \fref{wall_F0} are the {\bf concave} and
{\bf convex cusps} at the endpoints of  
apparently linear intervals. In our notation (cf.~figure), the bend at
$\simeq 0.2$ is a convex cusp while the one at $\simeq 0.3$ is a
concave one.
We will explain now their origin and give
analytical expressions for their
positions. 

As we will illustrate with examples, this kind of structure appears at
those values of the couplings at which 
a transition between different cascades occurs. A semi-quantitave
measure of how different two cascades are  
is given by the number of steps $m$ that they share in common. In this
sense, if a given cascade A shares $m_1$ 
steps with cascade B and $m_2$ with cascade C, with $m_1>m_2$, we say
that A is closer to B than to C. The  
general principle is that the closer the cascades between which a
transition occurs at a given initial  
coupling, the smaller the corresponding feature in the position of
duality wall versus coupling curve is. 

It is important to remember what the physical meaning of our
computations is. 
Numbering cascade steps increasing towards the UV and identifying the 
values of the initial couplings 
are just a simple way to handle the process of
reconstructing a duality cascade. This  
cascade represents a traditional RG flow in the IR direction, in which
Seiberg duality is used to switch  
to alternative descriptions of the theory beyond infinite coupling. At
some stage of this flow in the IR the 
model in \fref{f:F0-2} appears, 
with couplings given precisely by what we called
initial conditions. Thus, two cascades 
that share a large number of steps $m$ in common, correspond to two RG
flows initiated at different  
theories with large gauge groups and number of bifundamental fields in
the UV that converge at some point,  
sharing the last $m$ steps prior to reaching the model in
\fref{f:F0-2}. Due to the fact that a
duality wall exists,  
the independent flows before convergence of the cascades take place in
a very small range of energies. 

We now investigate the convex and concave cusps of the curve. 
Our approach consists of identifying what happens to the cascades at
those special points, and then  
computing the corresponding values of the initial couplings
analytically. Let us first consider  
the {\em concave cusps}. The $m$-th concave cusp corresponds to the
transition from node 3 being dualized  
at step $m+1$ to it being dualized at step $m+2$. The cascades at both
sides of the $m$-th concave cusp  
share the first $m$ steps and are of the form

\beq
\begin{array}{l}
2121 \ldots a 3 \\
\underbrace{2121 \ldots a}_{m} b3
\end{array}
\label{cascade_concave_cusps}
\eeq
where $(a,b)=(1,2)$ for $m$ even and $(2,1)$ for $m$ odd. In this way,
concave cusps fit in our general  
discussion of transitions between cascades, and we see that cusps
become smaller as $m$ is increased. The  
values of $x_3(0)$ that correspond to the concave cusps are obtained
by setting $x_3(k)=0$ in \eref{x3_step_k} and \eref{x3b1}
for $k \geq 2$, i.e.
\beq
x^{conc}_3(k)={2 \over 9} \sum_{i=2}^k {(7 i+4) \over i^2 (i+1)} \ \ \
\  k \geq 2 \ . 
\label{position_concave_cusps}
\eeq
 From \eref{position_concave_cusps}, the first concave cusps are
located at $x_3(0)$ equal to 
\beq
{1 \over 3}, \ \ \ {79 \over 162}, \ \ \ {467 \over 810}, \ \ \ {2569
\over 4050}, \ \ \ {19133 \over 28350} \ ,  \ldots
\eeq
in complete agreement with the numerical values of Figures
\ref{wall_F0} and \ref{dwall_F0}.

Let us move on and study the convex cusps in \fref{wall_F0}. In
analogy with \eref{cascade_concave_cusps}, we claim 
that the $m$th convex cusp corresponds to cascades switching between

\beq
\begin{array}{l}
2121 \ldots a3a \\
\underbrace{2121 \ldots a}_{m} 3b
\end{array}
\label{cascade_convex_cusps}
\eeq
with $(a,b)=(1,2)$ for $m$ even and $(2,1)$ for $m$ odd. In order to
check whether the proposal in \eref{cascade_convex_cusps} 
is correct, we proceed to compute the positions for the cusps that it
predicts. The calculation is similar to the one in 
\sref{sectionRGflow} and we only quote its result here

\beq
x^{conv}_3(k)={(4+7k)(10+49k+50k^2+14k^3)\over 9 k^2 (1+k)^2 (3+22k+14k^2)}+
{2 \over 9} \sum_{i=2}^{k-1} {(7 i+4) \over i^2 (i+1)} \ , \qquad
  k \geq 2 \ .
\label{position_convex_cusps}
\eeq

Equation \eref{position_concave_cusps} determines the following
positions for the first convex cusps 

\beq
{70 \over 309}, \ \ \ {21773 \over 50544}, \ \ \ {76733 \over 141750},
\ \ \ {457831 \over 750060}, \ \ \ {83386559 \over 126809550} \ ,
\ldots 
\eeq
which are in perfect
accordance with Figures \ref{wall_F0} and \ref{dwall_F0}, whereby
validating \eref{cascade_convex_cusps}. 

\paragraph{The Fractal:}

Something fascinating happens when the duality wall curve is studied
in further detail. Although convex cusps appear as  
such when looking at the curve at a relatively small resolution as in
\fref{wall_F0}, an
infinite fractal series of concave and convex cusps  
blossoms when we zoom in further and further. 
As an example, we show in
\fref{F0_fractal_zoom} successive amplifications of the area around
the  
first convex cusp, indicating the dualization sequences associated to
each side of a given cusp. According to our previous discussion,  
this cusp is located at $x_3(0)=70/309$ and corresponds to the
transition between two cascades differing at the third step:  
$232 \ldots$ and $231 \ldots$. \fref{F0_fractal_zoom}.b zooms in. We
can appreciate that what originally seemed to be a convex  
cusp becomes a pair of convex cusps with a concave one in the
middle. Furthermore, the value of $x_3(0)$ given by  
\eref{position_convex_cusps} is in fact the one that corresponds to
this originally hidden concave cusp. The new convex cusps are of  
a higher order, corresponding to transitions between cascades at the
4th step. The first one in \fref{F0_fractal_zoom}.b corresponds  
to $2323 \ldots \rightarrow 2321 \ldots$ while the second one is
associated to $2312 \ldots \rightarrow 2313 \ldots$. We see in  
\fref{F0_fractal_zoom}.c how each of the convex cusps splits again into
two 5th order convex cusps with a concave one in between.

This procedure can be repeated indefinitely. We 
conclude that concave cusps
are fundamental, while an infinite self-similar structure  
that corresponds to increasingly closer cascades can be found by
expanding convex cusps.

\FIGURE[ht]{
  \epsfxsize = 10cm
  \centerline{\epsfbox{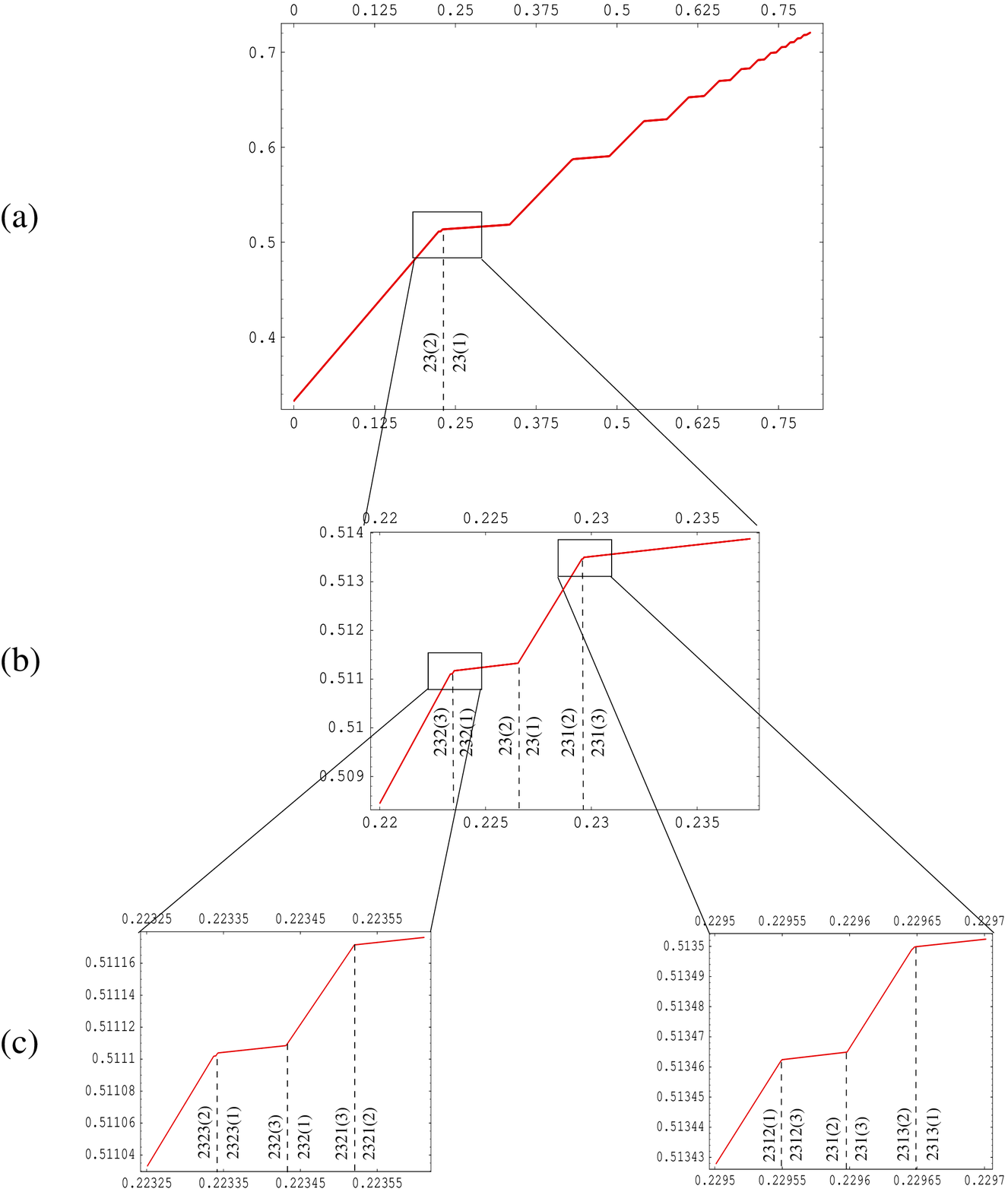}}
  \caption{{\sf Succesive amplifications of the regions around convex
  cusps show the self-similar nature of the curve for the  
                position of the wall versus $x_3(0)$. We show the
  first steps of the cascades at each side of the cusps, 
                indicating between parentheses the first dualizations
  that are different.}} 
  \label{F0_fractal_zoom}}


\section{RG Flows and Quasiperiodicity}

Having expounded in detail the analytic treatment of RG flows for the
zeroth Hirzebruch theory as well as their associated fractal behavior,
let us move on to see what novel features arise for more complicated
geometries. We recall the next simplest del Pezzo surface is the blow
up of $\IP^2$ at 1 point, the so-called $dP_1$. The gauge theory for
D3-brane probes on the cone over $dP_1$ was constructed via toric algorithms
in \cite{Feng:2000mi}. There are infinitely many quiver gauge theories
which are 
dual to this geometry. Their connections under Seiberg duality can be encoded 
in a duality tree. When D5-branes are included, the duality
tree becomes a representation of the possible paths followed by a cascading
RG flow. The tree for $dP_1$ appears in Figure 18 of \cite{Franco:2003ja}.
This tree contains isolated sets
of quivers with conformal ranks $r=(1,1,1,1)$, 
denoted toric islands in \cite{Franco:2003ja}. 
We will 
find quasiperiodicity of the gauge couplings for RG cascades 
among these 
islands.

\subsection{Initial Theory}

We are interested in studying the RG flow of a gauge theory corresponding
to $dP_1$.  For simplicity, let us choose one of the dual quivers with a 
relatively small number of bifundamentals.
Our quiver is described by the following
(we have also included the inverse matrix as a preparation to compute
the R-charges):
\be
\ba{l}\epsfxsize = 6cm \epsfbox{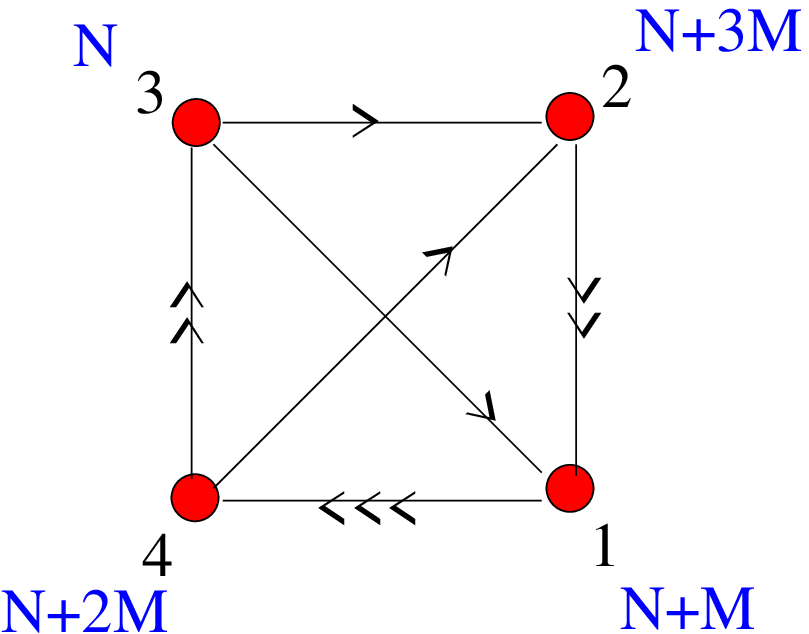}\ea
~~
S = \mat{
1 & -2 & -1 & 3 \cr
0 & 1 & -1 & -1 \cr
0 & 0 & 1 & -2 \cr
0 & 0 & 0 & 1 \cr
}
\ ,
\qquad
S^{-1} = \mat{
1 & 2 & 3 & 5 \cr
0 & 1 & 1 & 3 \cr
0 & 0 & 1 & 2 \cr
0 & 0 & 0 & 1 \cr
}
\ .
\ee
We start with a gauge theory with $N$ D3-branes
and $M$ D5-branes, $M \ll N$, corresponding to
gauge groups
\be
SU(N+M) \times SU(N+3M) \times SU(N) \times SU(N+2M) \ .
\ee
Chiral anomaly cancellation is satisfied since the 
D3-brane vector 
$r=(1,1,1,1)$ 
and the D5-brane vector $s=(1,3,0,2)$ are in the kernel of $S-S^T$. 
In fact, the kernel of $S-S^T$ is two dimensional, and these
are the only kinds of D-branes that are allowed.
The R-charges of the bifundamental fields 
at the conformal point are then, using \eref{R}, 
\begin{eqnarray}
R(X_{32}) &=& \frac{1}{4} \ , \nn\\
R(X_{21}) &=&  R(X_{43}) = \frac{1}{2} \ , \nn \\
R(X_{42}) &=& R(X_{31}) = R(X_{14}) = \frac{3}{4} \ .
\end{eqnarray}
As before, we assume the conformal 
R-charges get corrections only
at order $(M/N)^2$. Subsequently, using \eref{beta}
we calculate the one loop beta functions for the four
gauge groups to be
\beq
\label{betadP1}
\beta/M = (-15/4, 27/4, -27/4, 15/4) \ .
\eeq


\subsection{RG Flow}

As discussed above, we let the gauge couplings evolve according 
to the beta functions and we perform a Seiberg duality on
the gauge group factor whose coupling diverges first.  
Interestingly, a Seiberg duality on node 2 or 3 produces
the same quiver up to permutation (with the rank of the dualized
gauge group appropriately modified). On the other hand, Seiberg duality 
on nodes 1 or 4 produces a different quiver
with larger numbers of bifundamentals.  

In the next section, we will perform a numerical study of the possible 
flows. We will see how certain RG flows involve a single type of quiver
and periodically return to the starting point up to a change in the
number of D3-branes. These cases are the $dP_1$ analogues of the KS 
cascade. We will also discover other more intricate flows with
a beautiful structure, depending on the initial conditions.  
We will describe the KS type flows analytically in \sref{s:analytical_evolution}.


\subsubsection{Poincar\'{e} Orbits}

Let us explore the two-dimensional space of initial couplings 
$(c-x_3(0)-x_4(0),0,x_3(0),x_4(0))$, where $c$
is some constant that fixes the overall normalization. Next,
choose some initial value for the pair $(x_3(0),x_4(0))$ and 
evolve the cascade for a large number of steps. An interesting 
way of visualizing these flows is the following. We keep all the 
values of $(x_3,x_4)$ which are both non-zero, i.e., when either 
node 1 or 2 but neither node 3 nor 4 is dualized. A subsequent 
scatter plot can be made for these values, and is presented in 
\fref{f:poincare} for various choices of initial conditions, which
are identified by different colors.

\FIGURE[ht]{
\centerline{
\bt{cc}
(a)
\epsfxsize = 7.5cm
\epsfbox{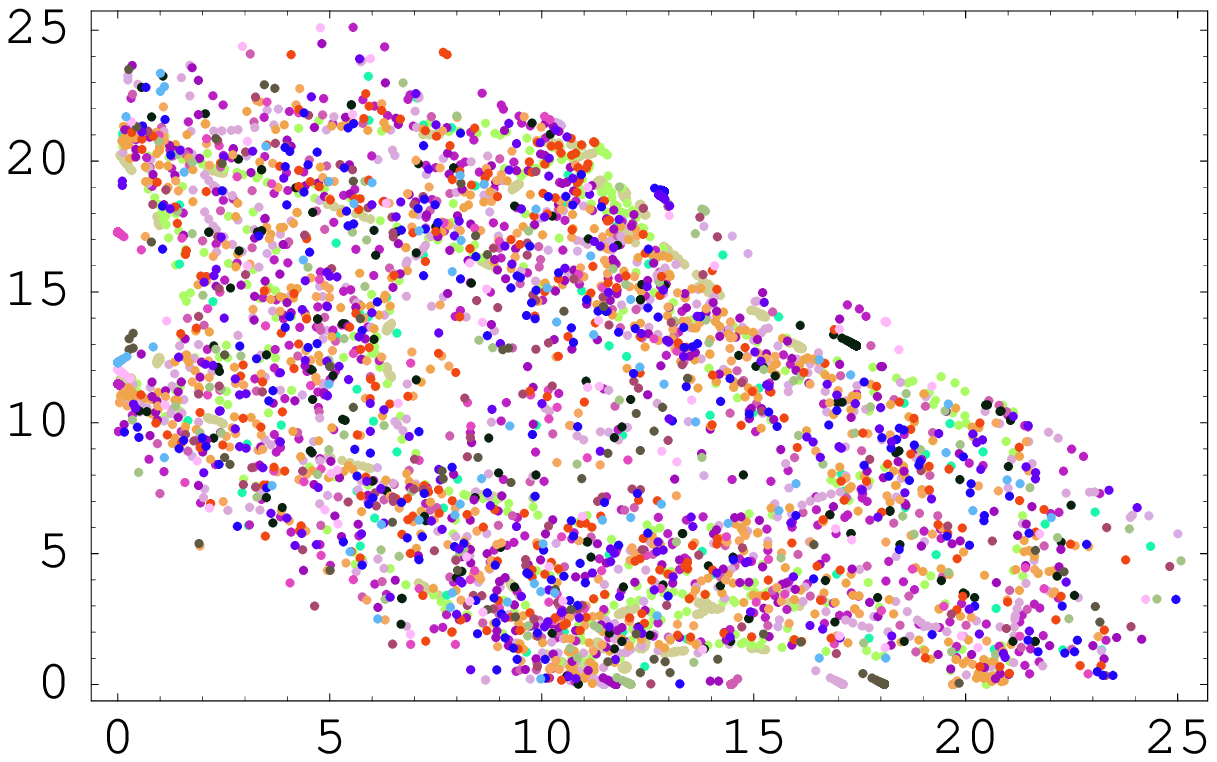}
&
(b)
\epsfxsize = 7.5cm
\epsfbox{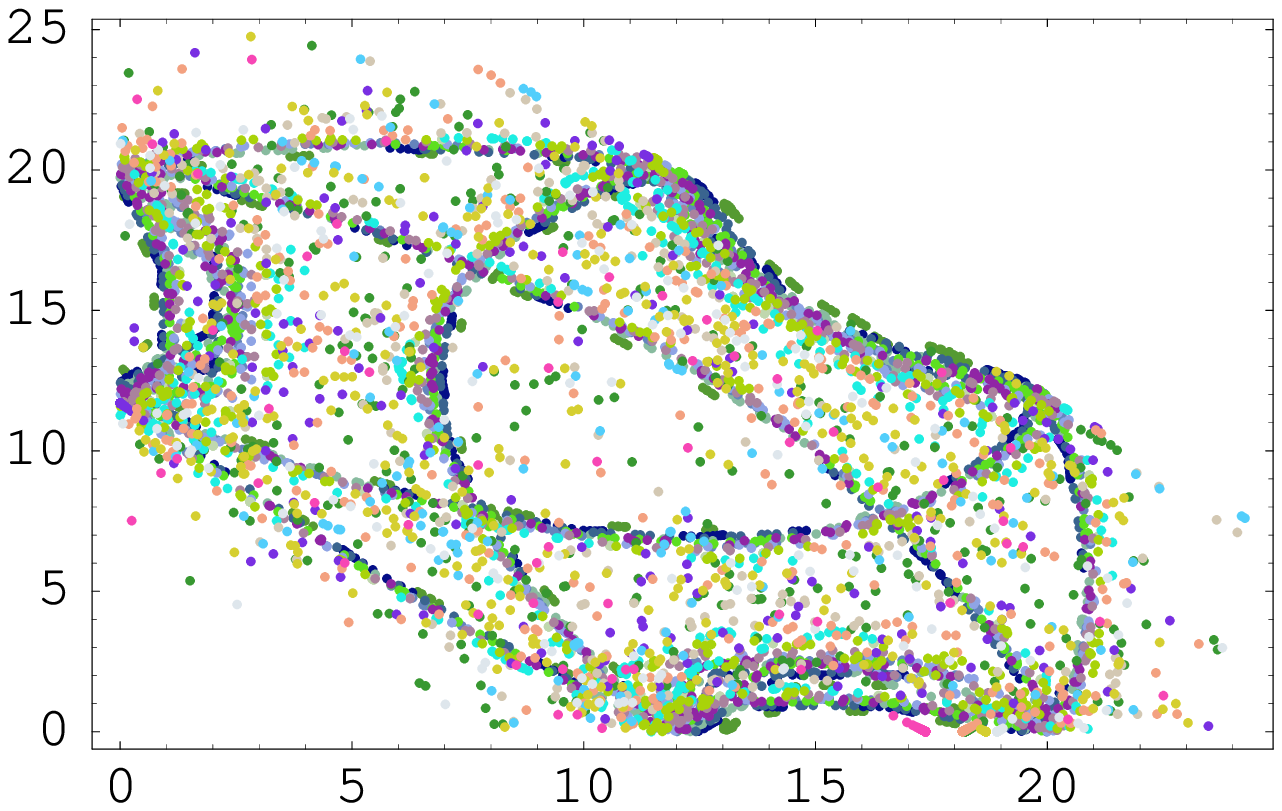}
\\
(c)
\epsfxsize = 7.5cm
\epsfbox{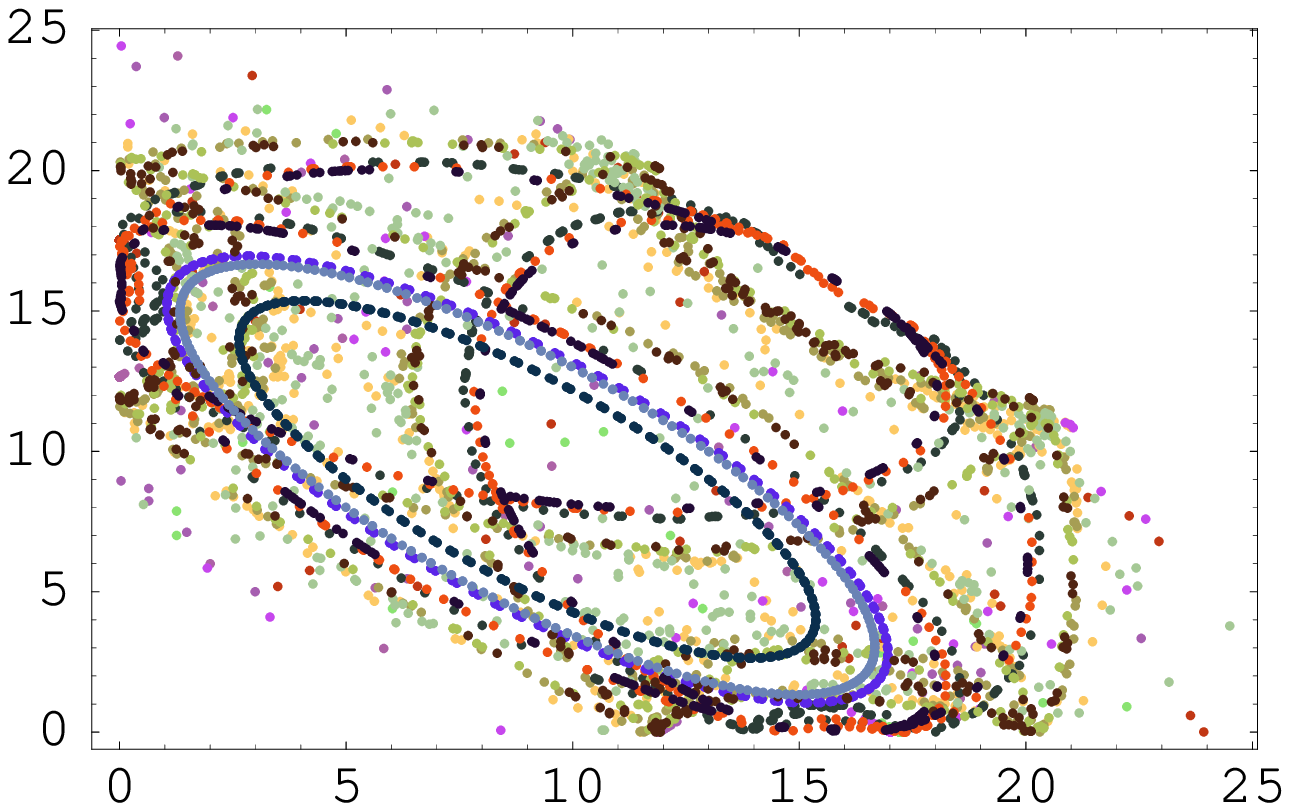}
&
(d)
\epsfxsize = 7.5cm
\epsfbox{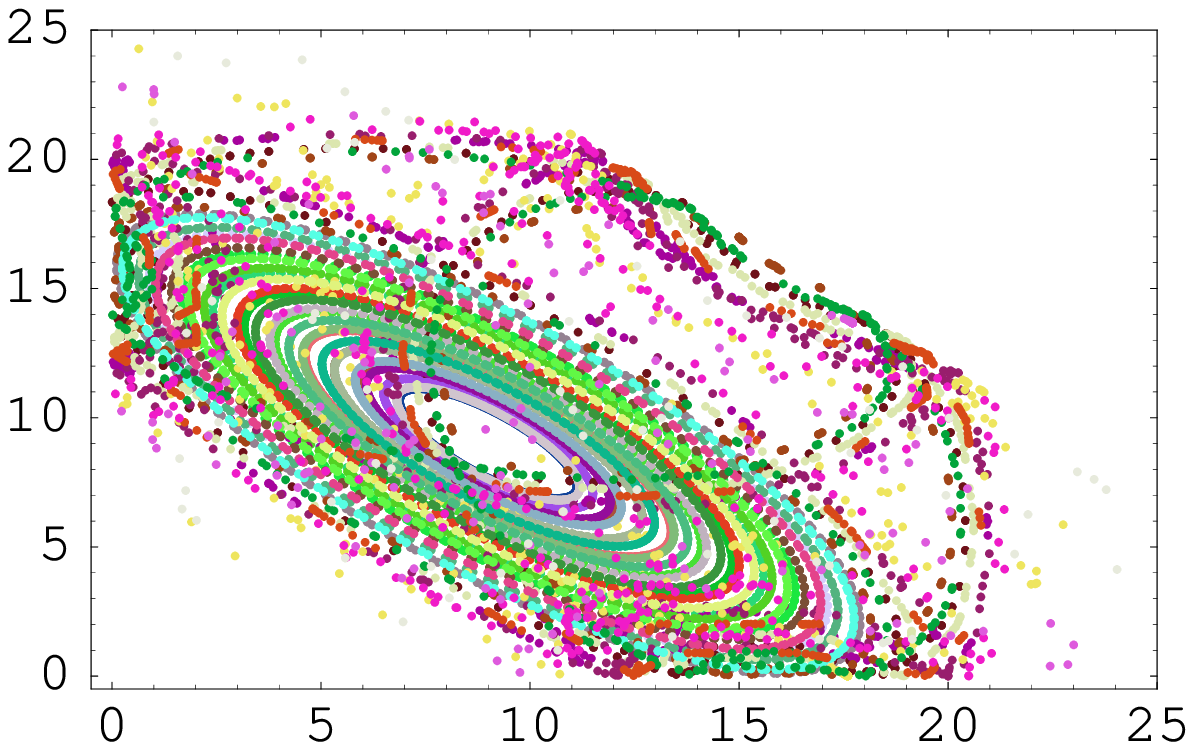}
\\
\et}
\caption{{\sf Scatter plot of $(x_3,x_4)$ that are non-zero during 800
dualization steps for the initial value
$(32-x_3(0)-x_4(0),0,x_3(0),x_4(0))$.  
In each plot, $(x_3(0), x_4(0))$ is allowed to range over
a rectangular region with lower left corner $L$,
upper right corner $R$, and minimum step size in 
the $x_3(0)$ and $x_4(0)$ directions equal to
$\delta_3$ and $\delta_4$ respectively.
(a) $L = (9, 15\frac78)$, $R = (10, 16\frac28)$,
$\vec \delta =(\frac14,\frac18)$;
(b) $L = (9, 15\frac38)$, $R = (10, 15\frac68)$,
$\vec \delta = (\frac14,\frac18)$;
(c) $L = (2, 6)$, $R= (5, 9)$, $\vec \delta = (1,1)$;
(d) $L = (7, 11)$, $R=(9, 17)$, $\vec \delta (1,\frac12)$. 
We use a different color for every
set of initial conditions.
}}
\label{f:poincare}}

We see different types of behavior according to the initial 
conditions. First, there are elliptical trajectories. They 
correspond to cascades that only involve $r=(1,1,1,1)$
quivers. In the language of \cite{Franco:2003ja}, the entire
RG flow takes place within a single toric island. The next section
will be devoted to a detailed study of this case. Other 
trajectories jump among three squashed ellipses. These cascades 
consist of both quivers with 
$r=(1,1,1,1)$ and $r=(2,1,1,1)$
(and its permutations) and correspond to hopping 
around the six toric islands. Finally, other flows have a 
diffuse scatter plot, and correspond to cascades that
travel to quivers with arbitrarily large gauge groups.
Outside the stable elliptical orbits, numerically we
find sensitive dependence on the initial conditions.

The scatter plots are reminiscent of the 
Poincar\'{e} surface-of-section (SoS) plots used in 
the study of chaotic dynamics.
We recall
that a Poincar\'{e} SoS is a surface in phase space which cuts the
trajectory of a system. 
If the trajectory is periodic or quasiperiodic,
the accumulation of intersection points where the trajectory
cuts the surface often produces cycles.
In our case, instead of phase space,
the RG cascade is a trajectory inside the
space of couplings, which we recall from \sref{s:simp} to be
a glued set of tetrahedra.  The ellipses we observe are sections
thereof.
In the above plots, we have actually superimposed different
surfaces, $x_2=0$ and $x_1=0$, but a symmetry has 
kept the picture from getting muddled. 



\subsubsection{Analytical Evolution}

\label{s:analytical_evolution}

Let us follow the RG flow analytically through several Seiberg dualities. 
We will focus on a particular sequence of dualities which repeats the 
sequence of dualizations on nodes 3, 1, 4 and 2. We will check later 
that this is indeed a consistent cascade that takes place once the initial
conditions are chosen appropriately. This set of dualizations never changes the 
quiver, but merely amounts to a permutation of the nodes after each step. 
 Furthermore, after four steps the cascade returns to the original quiver, 
with the same ordering of the nodes, but with the ranks changed as: 
$N_i \to N_i+4M$.

Now we are ready to try to understand the regime of initial
conditions which will allow for such a flow.
Let the initial inverse gauge couplings be
$x = (x_1, x_2, x_3, x_4)$ and set $M=1$.
The change in couplings from four steps of Seiberg dualities $(3142)$ 
can be cast as a linear
operation sending $x \to \cM x$ where $\cM$ is a $4\times 4$
matrix:
\beq\label{Mcoup}
\cM =
\left(
\begin{array}{cccc}
-55/729 & 5/9 & 6440/6561 & 56/81 \\
0 & 0 & 0 & 0 \\
154/729 & -5/9 & -1265/6561 & 70/81 \\
70/81 & 1 & 154/729 & -5/9 
\end{array}
\right) \ .
\eeq
In particular, $\cM$ has eigenvectors
\be
\lambda = 0, 1, \frac{-5983 \pm 1904 i \sqrt{2}}{6561} \ .
\ee
The zero eigenvalue has eigenvector $v_0 = (-5,9,-9,5)$, which 
can be used to set $x_2 = 0$.
The eigenvalue $\lambda = 1$ has eigenvector $v_1 = (14,0,9,9)$, and 
corresponds to a fixed point of the flow. If $x = v_1$, then the couplings 
will remain unchanged after a sequence of four Seiberg dualities. The 
normalization of this vector is the same one that was used in \fref{f:poincare},
where we can verify that the center of the ellipses is accordingly located at
$(x_3,x_4)=(9,9)$.
 Finally, the two complex eigenvalues, which we note to have unit
modulus, and henceforth define to be
\beq
\lambda_\pm := e^{\pm i\theta} \ ,
\eeq 
correspond to eigenvectors
\be
v_\pm := \left( 
\frac{2}{3} e^{\pm i \alpha}, 0, e^{\pm i \beta}, 1
\right) 
\ee
where
\be
e^{i\alpha} = 
-\frac{1}{3} + \frac{2i\sqrt{2}}{3} \; ; \; \; \;
e^{i\beta} =  -\frac{7}{9} - \frac{4i\sqrt{2}}{9} \ .
\ee

Let us take our initial couplings to be 
$x = a v_1 + c v_+ + c^* v_-$ for coefficients $a$ and $c$.  
After a large
number of Seiberg dualities, the couplings become, by \eref{Mcoup},
\be
x(n) = \cM^n x = av_1 + c \lambda^n_+ v_+ + c^* \lambda^n_- v_- \ .
\ee
The components of $x(n)$ are straightforwardly obtained, using the
above expressions for the various eigenvalues, as
\begin{eqnarray}
x_1 &=& 14 a + \frac{4}{3} |c| 
\cos(n\theta + \alpha + \delta) \nonumber \\
x_2 &=& 0 \nn \\
x_3 &=& 9a + 2|c| \cos (n\theta + \beta + \delta) \nonumber \\
x_4 &=& 9a + 2|c| \cos(n\theta + \delta) \label{xis}
\end{eqnarray}
where we have set $c = |c| \exp(i \delta)$.
We see that indeed, $(x_3,x_4)$ give rise to the parametric equation
for an ellipse with respect to the parameter $n$, in accord
with the scatter plots (c) and (d)
in \fref{f:poincare}. However, we must ask when is the above analysis
applicable, i.e., when is our dualization sequence actually 
the sequence followed by the RG flow.
Certainly
a necessary condition is that the couplings
$x_1$, $x_3$, and $x_4$ remain greater than zero
during the flow.  Thus, we see that $|c| < 9a/2$
with $a>0$. Indeed, under the condition $|c| < 9a/2$, an elliptical disk in 
the coupling plane $x_2=0$ is traced.  The boundary of the disk is
an ellipse tangent to the $x_3=0$ and $x_4=0$ axes
at the points $x/a = (16,0,0,16)$ and $x/a=(16,0,16,0)$.
This condition also appears to be sufficient,
as the numerics bear out.  Initial conditions violating this condition
will not generate ellipses, as demonstrated by plots (a) and (b).

Though one might worry, there is in fact
no contradiction between this periodic behavior and
the expectation that under RG flow, there will be fewer degrees of freedom in
the IR than in the UV.  
This expectation has been encoded more precisely in the
so-called $a$-conjecture (see for example \cite{IW} for a recent discussion).  
One can associate to any four-dimensional
conformal theory a central charge denoted $a$ which can be interpreted as a measure
of the number of degrees of freedom in the theory.  According to the $a$-conjecture,
given UV and IR conformal fixed points, $a_{UV}> a_{IR}$.
Now for our field theory analysis to be valid, our gauge theories should never be
very far away from conformality, where this distance is
measured by the ${\mathcal O}(M/N)$ corrections.  One expects therefore
that $a$ can be loosely defined at any point in the RG cascade and moreover
that $a$ should be non-increasing as we move into the IR.  
Recall that $a \sim \sum_\psi R(\psi)^3$
where the sum runs over the R-charges of all the fermions in the 
theory \cite{anselmi}.  
 From the structure of these quiver theories, one sees that 
$a \sim N^2$ and moreover after a sequence of four dualities
for the $dP_1$ flow above, $N \to N-4M$.
Thus $a$ is indeed decreasing as we move into the IR despite the 
periodic behavior of the gauge couplings.

\paragraph{Increments in Energy Scale}

One final question we can answer here is how does
the RG scale grow along the flow.
After a sequence of four
Seiberg dualities,
the RG scale changes by
\be
\Delta t(n) = \frac{4}{27} 
\left(\frac{106}{81} x_1(n) + x_2(n) + \frac{1108}{729} x_3(n) + \frac{4}{9} x_4(n)
\right)
\ .
\ee
In deriving this formula, we have had to look at 
the effect on the couplings of each of the four
Seiberg dualities individually.  The process is very similar 
to the calculations discussed in \sref{sectionRGflow} and we will not
repeat the details here.
Using the results (\ref{xis}), one finds that
\be
\Delta t (n) = \frac{16}{3} a + \frac{2^5 \cdot 7}{3^6} |c| 
\cos(n\theta + \delta + \gamma)
\ee
where 
\be
e^{i\gamma} = 
-\frac{241}{243} - \frac{22i \sqrt{2}}{243} \ .
\ee
Note that $\Delta t >0$, but that $t$ will have oscillations
on top from the cosine:
\be
t(n) = \frac{16}{3}an + \frac{2^5 \cdot 7}{3^6} |c| \frac{\cos(\delta + \gamma +
n\theta/2) \sin((n+1)\theta/2)}{\sin(\theta/2)} \ . 
\label{analytic_scale}
\ee

The previous approach can be applied to periodic KS type cascades
associated to other geometries. In the general case, as in the 
$F_0$ example of \fref{f:F0-1}, more than one quiver
can be involved in a period.

\newpage

\section{Supergravity Solutions for del Pezzo Flows}

In the above, we have discussed in detail the 
RG flows for some del Pezzo gauge theories from a purely
field-theoretic
point of view. This is only half of the story according to the AdS/CFT
Correspondence. It is important to find type IIB 
supergravity solutions that are dual to these field theory flows.
As already emphasized \cite{KS}, the main reason to trust that 
Seiberg duality cascades occur for the KS solution is not the field
theory analysis but that it is reproduced by a well behaved supergravity 
solution. The purpose of this 
section is to investigate these dual solutions.

Surprisingly, even without a metric for the del Pezzos, we can
demonstrate the existence of and almost completely characterize some
of their supergravity solutions.  
The solutions we find are analogous to the
Klebanov-Tseytlin (KT) solution \cite{KT} for the conifold.  Recall
that the KS solution is well behaved everywhere 
and asymptotes to the KT solution in the ultraviolet (large radius).
The KT solution, on the 
other hand, is built not from the warped deformed conifold but from
the conifold itself 
and thus has a singularity in the infrared (small radius).

\subsection{Self-Dual (2,1) Solutions}

To put these type IIB 
SuGRA solutions in historical context, 
note that they are closely related to
a solution found by Becker and Becker \cite{BB} for
M-theory compactified on a Calabi-Yau four-fold with 
four-form flux.  One takes the four-fold to be
a three-fold $\bX$ times $T^2$ and then $T$-dualizes
on the torus, as was done in \cite{GVW,DRS}.
The crucial point here is that the resulting 
complexified three-form
flux has to be imaginary self-dual and a harmonic
representative of $H^{2,1}(\bX)$
to preserve supersymmetry. 
Gra\~na and Polchinski \cite{GP} and also Gubser \cite{G} 
later noticed that the KT and KS supergravity solutions
were examples of these self-dual (2,1) type IIB 
solutions.  (Indeed, the authors of \cite{KS} also mention
that their complexified three-form is of type (2,1).)

Let us briefly review the work of Gra\~na and Polchinski.
The construction begins with a warped product of $\IR^{3,1}$ and a
Calabi-Yau three-fold $\bX$:
\be
ds^2 = Z^{-1/2} \eta_{\mu\nu} dx^\mu dx^\nu + Z^{1/2} ds^2_\bX \ ,
\ee
where the warp factor $Z=Z(p)$, $p \in \bX$, depends on only the
Calabi-Yau coordinates. 
We are interested in the case where $\bX$ is the total space of the
complex line bundle ${\mathcal O}(-K)$ over the del Pezzo $dP_n$. 
Here $K$ is the canonical class. The manifold $\bX$ is noncompact. 

There exists a class of
supersymmetric solutions with nontrivial flux
\be
G_3 = F_3 - \frac{i}{g_s} H_3
\ee
where $F_3=dC_2$ is the RR three-form field strength and $H_3=dB_2$
the NSNS three-form. 
To find a supergravity solution, the complex field strength $G_3$ must
satisfy several conditions: $G_3$ must 
\begin{enumerate}
\item be supported only in $\bX$;
\item be imaginary self-dual with respect to the Hodge star on $\bX$,
	i.e., $\star_\bX G_3 = iG_3$; 
\item have signature $(2,1)$ with respect to the complex structure on
$\bX$; and finally,
\item be harmonic.
\end{enumerate}
If these conditions are met, a supergravity solution exists such that the 
RR field strength $F_5$ obeys
\be\label{F5}
dF_5 = - F_3 \wedge H_3 \ ,
\ee
and the warp factor satisfies
\be\label{warp1}
(\nabla_\bX^2 Z) \vol(\bX) = g_s F_3 \wedge H_3 \ ,
\ee
where $\vol(\bX)$ is the volume form on $\bX$.
In particular, $\vol(\bX) = r^5 dr \wedge \vol(\Y)$
where $\Y$ is the (5 real-dimensional) level surface of the cone $\bX$.
The axion vanishes and the dilaton is constant: $e^\phi=g_s$.

\subsection{(2,1) Solutions for the del Pezzo}
Let us construct such a $G_3$ for the del Pezzos.  
As a first step, we construct the metric on $\bX$.  
Let $h_{a\bar b}$ be a K\"{a}hler-Einstein
metric on $dP_n$
such that $R_{a\bar b} = 6h_{a\bar b}$. Indeed, we
only know of the existence of and not the explicit form\footnote{Such a
metric is known not to exist for $dP_1$ and $dP_2$. 
See for example \cite{Tian}.} of $h_{a\bar b}$.
We want to consider the case where $\bX$ is a cone over $dP_n$.  In
this case, the metric on $\bX$ can be written \cite{BH, BHK} as
\be
ds_\bX^2 = dr^2 + r^2 \eta^2
+ r^2 h_{a \bar b} dz^a d{\bar z}^{\bar b} \  ,
\ee  
where $\eta = \left(\frac{1}{3} d\psi + \sigma \right)$.
The one-form $\sigma$ must satisfy $d\sigma = 2 \omega$ where $\omega$
is the K\"{a}hler form on $dP_n$ and $0 \leq \psi < 2\pi$ is the
coordinate on the circle bundle over $dP_n$.

Next, we describe a basis of self-dual and anti-self-dual harmonic 
forms on $dP_n$.\footnote{We would like to thank Mark Stern and 
James M\raise 2pt \hbox{c}Kernan for the following argument.}
We begin with the K\"{a}hler form $\omega$.
Locally, $dP_n$ looks like $\IC^2$ and 
$2\omega \sim dz^1 \wedge d\bar z^ {\bar 1} + dz^2 \wedge d\bar
z^{\bar 2}$.  Thus locally, it is easy to see that $\omega$ is
self-dual under the operation of the Hodge star. 
Because the Hodge star is a local operator, $\omega$ must be self-dual
everywhere. Now, recall our $dP_n$ are Einstein.  Thus 
\be\label{omega}
\omega = 6 iR_{a\bar b} dz^a \wedge d\bar z^{\bar b} = 6 i\partial
\bar \partial \ln \sqrt{\det{h}}
\ .
\ee 
Clearly $d\omega = (\partial + \bar \partial ) \omega = 0$ whence
$\omega$ must be 
closed.  It follows that $\omega$ is a self-dual harmonic form on
$dP_n$.

There exists a cup product (bilinear form) $Q$ on $H^{1,1}(dP_n)$
defined as
\be\label{cup}
Q(\phi, \xi) = \int_{dP_n} \phi \wedge \xi \ , \qquad
\phi, \xi \in H^{1,1}(dP_n) \ .
\ee      
The Hodge Index Theorem states that $Q$ has signature $(+,-,\ldots,
-)$. For $dP_n$, $h^{2,0}=0$ while $h^{1,1} = n+1$, there being
$n$ other harmonic $(1,1)$  forms
on $dP_n$ in addition to $\omega$.  We denote these harmonic forms
as $\phi_I$, $I=1,\ldots, n$.  Let us pick a basis for $Q$ such that 
\beq\label{phiI}
\phi_I \wedge \omega = 0 \ .  
\eeq
From the above discussion of $\omega$
one sees that
\be
0 < \int \omega \wedge \star \omega = \int \omega \wedge \omega
\ee
where the inequality follows from the definition of the Hodge star 
and the equality from the fact that $\omega$ is self-dual.
Hence the $\phi_I$ span a vector space $V$ where
$Q$ has purely negative signature.

Recall that the Hodge star in two complex dimensions squares to one:
$\star \star \phi = \phi$.  Thus we can diagonalize $\star$ on $V$
such that $\star \phi_I = \pm \phi_I$.  However, if $\star \phi_I =
\phi_I$, then
one would find $\int \phi_I \wedge \phi_I > 0$, in contradiction to
the fact that $Q$ has purely negative signature on $V$.  We conclude
that the $\phi_I$ must all be purely anti-self-dual, $\star \phi_I =
-\phi_I$. 

With these preliminaries,
it is now straightforward to construct $G_3$.  We let
\be\label{H3}
F_3 = \sum_{I=1}^k a^I\, \eta \wedge \phi_I \ , \qquad
H_3 = \sum_{I=1}^k a^I\, g_s \frac{dr}{r} \wedge \phi_I  \ ,
\ee
for expansion coefficients $a^I$.   Hence,
\be
G_3 = \sum_{I=1}^k a^I (\eta - i \frac{dr}{r}) \wedge \phi_I \ .
\ee
This is a solution because by construction, $G_3$ is
harmonic and is supported on $\bX$ so conditions (1) and (4) are met.
Moreover, $(dr / r + i \eta)$ is 
a holomorphic one-form on $\bX$. Therefore, $G_3$ must have signature 
$(2,1)$ because $\phi$ is a $(1,1)$ form. Furthermore,
it is easy to check that 
$\star_\bX G_3 = i G_3$.  Thus, conditions (2) and (3) are also
met.

\paragraph{D5-Branes}
The number of D5-branes in this SUGRA solution is given by the 
Dirac quantization condition on the RR flux. More precisely, we
have an integrality condition on the integral of $F_3$ over
compact three-cycles in the level surface $\Y$ of the cone
$\bX$. Given a basis $\calh^J$ ($J=1,\ldots n$) of 
such cycles, we impose that writing
\be
\int_{\calh^J} F_3 = 4 \pi^2 \alpha' M^J \ ,
\label{d5quant}
\ee
must give integer $M^J$. From the construction of $\Y$, it follows 
that ${\mathcal H}^J$ will be some circle bundle over a curve $D^J 
\subset dP_n$ while the circumference of the circle is $2\pi/3$. 
Subsequently, equation (\ref{d5quant}) reduces to
\be
\sum_I a^I \int_{D^J} \phi_I = 6 \pi \alpha' M^J \ .
\label{Dj}
\ee
To understand the curve $D^J$, we take a closer look at the divisors 
that correspond to elements of $H^{1,1}(dP_n)$.  Because $dP_n$ is
${\mathbb P}^2$ blown up at $n$ points, there will be a divisor $H$ 
corresponding to the hyperplane in ${\mathbb P}^2$ and exceptional 
divisors $E_i$ ($i=1,\ldots,n$) for each of the blow ups.  Essentially 
because two lines intersect at a point, $Q(H,H) = H\cdot H = 1$. 
From the blow-up construction, we also know that $Q(E_i, E_j) = 
E_i \cdot E_j = - \delta_{ij}$.  Finally, $E_i \cdot H = 0$ because
the blow-ups are at general position.  We see
explicitly that $Q$ has signature $(+, -, -, \ldots ,-)$.  
From Poincar\'e duality, there is a one-to-one 
map from the differential forms $\omega$ and $\phi_I$ to the divisors
$H$ and $E_i$, which we now explore.

The first chern class of ${\mathbb P}^2$ is $c_1({\mathbb P}^2) = 3H$.
By the adjunction formula, 
it follows that $c_1(dP_n) = 3H - \sum\limits_{j=1}^n E_j$.
Locally, the first
chern class can be expressed in terms of the Ricci tensor, 
\be
c_1(dP_n) = i \frac{R_{a\bar b}}{2\pi} dz^a \wedge d\bar z^{\bar b} \ ,
\ee  
and then from the Einstein condition \eref{omega}, we find that
\be
\omega = \frac{\pi}{3} c_1(dP_n) \ .
\ee  
Thus, by \eref{phiI}, the $\phi_I$ must be orthogonal to $c_1(dP_n)$.  
This orthogonality condition has an astonishingly beautiful (and well
known) consequence. The orthogonal complement of $3H - \sum_j E_j$ 
is the weight lattice of the corresponding exceptional Lie group 
${\mathcal E}_n$. In this language the $\phi_I$ must lie in this 
weight lattice.  

We now return to the question, what are the curves $D^J$ in the 
integral (\ref{Dj})?  The problems we need to worry about in
defining the $D^J$ are essentially the same problems we need to 
worry about in trying to quantize the flux in a far simpler system, 
that of a collection of point electric charges in three dimensions.  
In drawing a sphere (or perhaps some shape with more complicated 
topology) around each charge, we want to make sure that the sphere 
wraps around the selected charge exactly once and no other charges.  

 For the $dP_n$, this condition translates into the requirement that
\be
\int_{D^J} \phi_I = \int_{dP_n} \phi_I \wedge c_1(D^J) = \delta^J_I \ .
\label{orth}
\ee
Because $\phi_I \wedge \omega = 0$, only the component of
$c_1(D^J)$ orthogonal to $c_1(dP_n)$ need be defined.  
Let us choose $c_1(D^J) \wedge \omega =0$.  
To avoid surrounding charges more than once, we need to
make the $D^J$ ``as small as possible''.  Thus we choose the $c_1(D^J)$
to be the generators of the weight lattice.
The condition (\ref{orth}) then implies that the $\phi_I$ generate the
root lattice. For example, for $dP_3$, we could choose
$\phi_1 = E_1 - E_2$, $\phi_2 = E_2 - E_3$, and $\phi_3 = H - E_1 -
E_2 - E_3$. 
Indeed, the bilinear form \eref{cup} can be written in the basis
\beq\label{cupCartan}
\int_{dP_n} \phi_I \wedge \phi_J = -A_{IJ}
\eeq
where
$A_{IJ}$ is the Cartan matrix for the ${\mathcal E}_n$ root lattice.

Finally, using \eref{H3}, \eref{Dj} and 
(\ref{orth}), we can normalize $F_3$ and $H_3$, giving
us
\beq\label{CjMj}
a^J = 6 \pi \alpha' M^J \ ;
\eeq
hence the number $M^J$ of D5-branes is fixed in our SUGRA solutions.
From a perturbative point of view, we can think of this SUGRA solution
as arising from the back reaction of D5-branes wrapped around vanishing
curves $C_I$ of $\bX$, which are the Poincar\'e duals of the
$\phi_I$. This follows from the definition $dF_3 = \sum a^I 
d(\eta\wedge\phi_I) = \sum a^I \delta_{C_I}$.

\paragraph{D3-branes}
Having discussed some detailed algebraic geometry for the $dP_n$,
we are now ready to quantize the number $N$ of D3-branes as well. The
condition reads, using \eref{F5},
\be\label{F5-quan}
\int_\Y F_5 = (4 \pi^2 \alpha')^2 N \ ,
\ee
where $F_5 = {\mathcal F} + \star_{10} {\mathcal F}$, 
and
\be
{\mathcal F} = 
\sum_{I,J} a^I a^J g_s \ln(r/r_0) \eta \wedge \phi_I \wedge \phi_J
 \ .
\ee
Therefore one finds, using \eref{cupCartan} and \eref{CjMj},
\be\label{F5-N}
N  = 
\frac{3}{2\pi} g_s \ln(r/r_0)
\sum_{I,J} M^I A_{IJ} M^J
\ee
for large $r$.  In other words, the number
of D3-branes grows logarithmically with the radius. 

\paragraph{Warp Factor}
Now, recalling from \cite{BH} that for $\Y = dP_n$,
\beq
\int_\Y  \vol(\Y) = \frac{\pi^3}{27}(9-n) \ ,
\eeq
we can use \eref{H3} and \eref{warp1}
to solve for the warp factor. The equation reads
\beq
\left[\frac{\partial^2}{\partial r^2} + 
\frac{5}{r} \frac{\partial}{\partial r}\right] Z(r) = 
\frac{(6 \pi \alpha' g_s)^2}{\Vol(\Y)}  \frac{2 \pi}{3 r}
\sum_{I,J}  M^I A_{IJ} M^J \ .
\eeq
This yields
\be
Z(r) = \frac{2 \cdot 3^4}{9-n} \alpha'^2 g_s^2 
\left(\frac{\ln(r/r_0)}{r^4} + \frac{1}{4r^4} \right) \sum_{I,J}  M^I
A_{IJ} M^J\ .
\ee

In short, we have found the analog of the Klebanov-Tseytlin solution, 
a solution that is perfectly well behaved at large 
radius but has a curvature singularity at small radius $Z(r_*)=0$.  
We envision that there is some similar warped deformed del Pezzo
solution which resolves the singularity, just as the warped deformed
conifold of the KS solution resolved the singularity of the KT solution.

%

\subsection{Gauge Couplings}

In order to move towards a comparison between SUGRA and gauge theory,
let us determine the gauge couplings on probe branes inserted into
the geometry we have discussed above. To begin with, let us study
D3-branes. Their gauge coupling is simply proportional to the
string coupling, $g_s$, which as we have seen is constant in the
self-dual (2,1) solutions. In gauge theory, this is expressed by
the fact that the sum of gauge couplings \eref{simprel} is independent 
of the scale.

We can also probe with D5-branes. Consider a D5-brane wrapped on
a curve $C_I\subset dP_n \subset\Y$ at a fixed radial position $r$ in $\bX$. 
We take $C_I$ to be the Poincar\'e dual of the harmonic two-form $\phi_I$. 
As is well-known (see, e.g., \cite{remarks}), the gauge
coupling on such a brane is related to the integral of the
NS 2-form around $C_I$ by
\beq
x_I = \frac{8\pi^2}{g_I^2} = -\frac1{2 \pi\alpha'g_s}\int_{C_I} B_2 \,.
\eeq
Thus, using the expression for $B_2$ by integrating $H_3$ from
\eref{H3}, as well as the value of 
$a^J$ from \eref{CjMj}, we find
\beq
x_I = - 3 \ln r \sum_J \int_{C_I} \phi_J M^J \ .
\eeq
This yields for the beta function
\beq
\beta_I = \frac{dx_I}{d\ln r} = -3 (C_I\cdot C_J) M^J \,,
\label{sugrabeta}
\eeq
where $C_I\cdot C_J =\int \phi_I\wedge\phi_J$ is the
{\it intersection pairing} of two-cycles in $dP_n$ and the sum
on $J$ is implied.

To compare this result with gauge theory, we first need to recall the fact
from section \ref{general} that a D5 brane wrapped around $C_I$ is 
associated with a certain combination of fractional branes that we 
have encoded in the vector $s_I= (s_I^i)$. Thus, the beta function
$\beta_I$ is related to the beta functions of the fractional branes
via
\beq
\beta_I = \sum_i s_I^i \beta_i \,,
\eeq
Inserting the expression for $\beta_i$ from \eref{beta}, we obtain
the gauge theory expression
\beq
\beta_I = 3 \sum_i  s^i_I s^i_J M^J + \frac32 \sum_{ij} s^i_I \Rt_{ij}
s^j_J M^J
\label{inter1}
\eeq
where $\Rt$ is given in \eref{Rt}. Let us now use the vanishing of
the beta function for the conformal theory (corresponding to putting
$d^i=r^i$ in \eref{beta1} and using \eref{beta}) 
to rewrite the first term as
\beq
\sum_i s^i_I s^i_J=
- \frac12 \sum_{ij} s^i_I s^i_J \Rt_{ij} \frac{r^j}{r^i} \ .
\eeq
Using the definition of $\Rt_{ij}$ in \eref{Rt}, we find the gauge theory
result
\begin{eqnarray}
\beta_I &=& \frac32 \sum_{ij} \Rt_{ij} \bigl( s^i_I s^j_J - s^i_I s^i_J 
\frac{r^j}{r^i}) M^J \nonumber \\
&=& 
\frac32 \sum_{ij} f_{ij} (R_{ij}-1) \bigl(
s^i_I s^j_J +  s^j_I s^i_J - 
s^i_I s^i_J \frac{r^j}{r^i} - s^j_I s^j_J \frac{r^i}{r^j}\bigr) M^J \,.
\label{gaugebeta}
\end{eqnarray}

To finish up and relate this long-winded expression to the intersection
pairing in \eref{sugrabeta}, we need to rely on certain results concerning
baryonic $U(1)$ charges in quiver gauge theories related to del Pezzos
\cite{MoPle,Intriligator:2003wr,HW,Herzog:2003wt}. 
First of all, these baryonic
$U(1)$ charges are in one-to-one correspondence 
with possible non-conformal
deformations. In formulas, one can write all baryonic $U(1)$ charges $Q_I$
as a sum
\beq
Q_I = \sum_{i} q^i_I Q_i
\eeq
where $Q_i$ is a charge associated with the nodes of the quiver and is equal 
to $+1$ for incoming arrows and $-1$ for outgoing arrows. In other words,
\beq
Q_I(X_{ij}) = q^j_I -q^i_I \ .
\eeq
For purposes of anomaly cancellation, these charges are
related to the null vectors $s^i_I$ of $\cali=S-S^t$ via 
\cite{Intriligator:2003wr,HW}
\beq
q^i_I r^i = s^i_I \,.
\eeq

It was then shown in \cite{HW} that the baryonic $U(1)$ charges $Q_I$ 
are also in one-to-one correspondence with curves $C_I$ in the del Pezzo 
orthogonal to the K\"ahler class. Moreover, it was shown in \cite{HW} 
that one could identify the intersection product of the curves $C_I$
as the cubic anomaly associated with the baryonic charges $Q_I$,
\be
C_I\cdot C_J = \frac12 {\rm tr} R Q_I Q_J \ .
\ee
Now let us relate this cubic anomaly to the beta functions.
\begin{eqnarray}
\left( {\rm tr} R Q_I Q_J \right) M^J
&=& 
M^J \sum_{i,j} f_{ij} (R_{ij}-1) Q_I(X_{ij}) Q_J(X_{ij}) r^i r^j \nonumber \\
&=& 
M^J \sum_{i,j} f_{ij} (R_{ij}-1) 
(q_I^j- q_I^i)(q_J^j-q_J^i) r^i r^j \nonumber \\
&=&
-M^J \sum_{i,j} f_{ij} (R_{ij}-1)
\bigl(
s^i_I s^j_J +  s^j_I s^i_J - 
s^i_I s^i_J \frac{r^j}{r^i} - s^j_I s^j_J \frac{r^i}{r^j}\bigr)
= -\frac{2}{3} \beta_I 
\,.
\nonumber
\end{eqnarray}
This expression, upon substituting into the SuGRA result
\eref{sugrabeta} gives the gauge theory result for the beta function
in \eref{gaugebeta}, whereby giving us the link we needed.

\subsection{Discussion}

On the one hand it is impressive that we can write down such a complicated
supergravity solution that encodes interesting field theoretic behavior 
without knowing the precise metric on the del Pezzos.  On the other, it
is a little disappointing that we have found no smoking gun for the
existence of duality walls from the supergravity perspective.

Let us consider the implications of this KT-like solution for del Pezzos.
Such a solution indicates that the dual del Pezzo field theory should
behave like the KS field theory.  In other words,
one expects a sequence of Seiberg dualities where as we move into
the UV, the number of
D3-branes gradually increases, the number of D5-branes remains fixed,
and no duality wall is reached.  We have seen such behavior for some
of the phases of the 
del Pezzos.  For example, the Model A/Model B flow of
\cite{Franco:2003ja} and 
the $dP_1$ flow considered here exhibit such behavior.  We expect
that such flows can probably be constructed for all del Pezzos.

Our supergravity solutions severely constrain possible
${\mathcal O}(M/N)$ corrections to the R-charges for
KS type flows.  In particular, both on the field theory side and
on the supergravity side, we saw that the sum of the beta functions
(\ref{simprel}) must vanish.  Additionally, we calculated the $n$
$\beta_I$ for $dP_n$ both in field theory and in supergravity
and saw that the two expressions agreed.  In total, we have
$n+1$ constraints on $n+3$ beta functions.  Thus, any corrections
to the beta functions for KS flows must lie in the remaining
two dimensional vector space.  (Note that for the original 
KT solution for the conifold,
the two constraints are enough eliminate any possible 
corrections to the two beta functions.)

We have also seen behavior vastly different
from KS type cascades.  For example, for
the $F_0$ surface, we saw duality walls.  Note also in this flow that
the number of D3-branes does not increase but is 
pinned by nodes three and four.  Presumably there is some other
supergravity solution which describes this flow.
One way of constructing a more general type of supergravity 
solution would be to try to construct $F_3$ with a dependence on
$dr$ or to start with a non-conical metric on $\bX$.

\section*{Acknowledgements}

We would like to thank Jerome Gauntlett, Ami Hanany, Aki Hashimoto,
James M\raise 3pt\hbox{c}Kernan, Rob Myers, 
Joe Polchinski, Mark Stern and Cumrum Vafa 
for useful discussions. This
research is funded in part by the CTP and the LNS
of MIT and by the department of Physics at UPenn. 
The research of S.~F. was supported in part by
U.S.~DOE Grant $\#$DE-FC02-94ER40818.
The research of Y.-H.~H. was supported in part by
U.S.~DOE Grant $\#$DE-FG02-95ER40893 as well as
an NSF Focused Research Grant
DMS0139799 for ``The Geometry of Superstrings''.
The research of C.~H. and J.~W. was supported in part by the 
NSF under
Grant No. PHY99-07949.
\newpage

\appendix

\section{Seiberg Duality and the Beta Function}

We show how to 
demonstrate that after Seiberg duality on node $i$ of a four-node, well split quiver,
the value of $\beta_i$ changes sign.  The proof makes extensive use of results
from \cite{H}.

Any well split, four-node quiver can be represented by the matrix,
\be
S = 
\left(
\begin{array}{cccc}
1 & a & b & c \\
0 & 1 & d & e \\
0 & 0 & 1 & f \\
0 & 0 & 0 & 1
\end{array}
\right)
\ee
where all the entries are integers, $b \geq 0$, $c \geq 0$, and $a$, $d$, $e$,
and $f$ are non-positive \cite{H}.  Note that one may have to
cyclically permute the ordering of the nodes to satisfy these sign
requirements.  In \cite{H} such a quiver was called $Ai$.
The cyclic permutations were labelled $Aii$, $Aiii$, and $Aiv$.
The conditions that $S$ be rank two
and that $\Tr S S^{-T} = 4$ put the following two constraints on 
the matrix entries:
\be
cd - be + af = 0 
\label{arel1}
\ee
and
\be
a^2 + b^2 + c^2 + d^2 + e^2 + f^2 - abd - ace - bcf - def + acdf = 0  .
\label{arel2}
\ee

In order to compute the beta functions, we also need expressions for the
ranks of the gauge groups at the conformal point.  Again from \cite{H},
we find that
\begin{eqnarray}
8r_1^2 &=& d^2 + e^2 + f^2 - def  \ ,\quad 
8r_2^2 = b^2 + c^2 + f^2  - bcf \ , \nonumber \\
8r_3^2 &=& a^2 +c^2 + e^2 - ace \ , \quad 
8r_4^2 = a^2 + b^2 + d^2 - abd \ , \nonumber \\
8r_1 r_2 &=& cdf - bd - ce \ , \quad 
8r_1 r_3 = ad - cf \ , \quad 
8r_1 r_4 = ae + bf - adf \ , \nonumber \\
8r_2 r_3 &=& acf - ab - ef \ , \quad 
8r_2 r_4 = -ac + fd \ , \quad 
8r_3 r_4 = acd - de - bc \ . 
\label{adranks}
\end{eqnarray}
Note that these values of the $r_i$ are independent of the signs
of the entries of $S$.

 Finally, we need to know how many D5-branes are present.
$S-S^T$ has only a two dimensional kernel.  We know that
$r$ is one element of the kernel.  From 
(\ref{arel1}), we can read off another, linearly independent element
$s = (0, f, -e, d)$.  The vectors $s$ and $r$
span the kernel, and we will assume we have one D5-brane, 
$M=1$, of the type $s$.

We now have enough to compute the beta functions using
(\ref{R}) and (\ref{beta}).  The expressions are messy, and we will
not reproduce them here.

To verify that the $\beta_i$ flips sign after Seiberg duality on 
node $i$, we have to see what happens to our quiver
after duality on nodes 1, 2, 3, and 4.

\paragraph{Node 1}

After Seiberg duality on node 1, the resulting quiver
can be described by the matrix
\be
S_1 = \left(
\begin{array}{cccc}
1 & a & d-ab & e-ac \\
0 & 1 & -b & -c \\
0 & 0 & 1 & f \\
0 & 0 & 0 & 1
\end{array}
\right)
\ee
and the vector $s$ is 
transformed into
$s' = (f, -af, -e, d)$.  The $r_i$
are still given by (\ref{adranks}) but
with the appropriate substitutions
indicated by $S_1$.

Note that Seiberg duality changes the ordering
of the nodes.  After duality, node 1 becomes
node 2.  Thus, we checked using
a computer algebra package
and the relations  (\ref{arel1}) and (\ref{arel2})
that $\beta_1 + \beta_2' = 0$.

\paragraph{Node 2}

After Seiberg duality on node 2, the resulting quiver
can be described by the matrix
\be
S_2 = \left(
\begin{array}{cccc}
1 & a & -d & -e \\
0 & 1 & b-ad & c-ea \\
0 & 0 & 1 & f \\
0 & 0 & 0 & 1
\end{array}
\right)
\ee
and the vector $s$ is 
transformed into
$s' = (-f, 0, -e, d)$.

After duality, node 2 becomes
node 1.  Thus, we checked 
that $\beta_2 + \beta_1' = 0$.

\paragraph{Node 3}

After Seiberg duality on node 3, the resulting quiver
can be described by the matrix
\be
S_3 = \left(
\begin{array}{cccc}
1 & -b & a-bd & c \\
0 & 1 & d & -f \\
0 & 0 & 1 & e-fd \\
0 & 0 & 0 & 1
\end{array}
\right)
\ee
and the vector $s$ is 
transformed into
$s' = (0, e-df, f, d)$.

After duality, node 3 becomes
node 2.  Thus, we checked 
that $\beta_3 + \beta_2' = 0$.

\paragraph{Node 4}

After Seiberg duality on node 4, the resulting quiver
can be described by the matrix
\be
S_4 = \left(
\begin{array}{cccc}
1 & e & f & c \\
0 & 1 & d & -a+ce \\
0 & 0 & 1 & -b+cf \\
0 & 0 & 0 & 1
\end{array}
\right)
\ee
and the vector $s$ is 
transformed into
$s' = (-d, f, -e, 0)$.

After duality, node 4 becomes
node 1.  Thus, we checked 
that $\beta_4 + \beta_1' = 0$.

\bibliographystyle{JHEP}

\end{document}